\documentclass[aps,twocolumn,superscriptaddress,footinbib,longbibliography]{revtex4-1}

\usepackage{longtable}
\usepackage{morefloats}
\usepackage[dvips]{graphicx}
\usepackage{color}
\usepackage{epsfig,graphicx,amsfonts,amsbsy}
\usepackage{amsmath,amsfonts,amsthm,amssymb}
\usepackage{appendix}
\usepackage{makeidx}
\usepackage{url}
\usepackage{verbatim}
\usepackage[bookmarksnumbered,pdfpagelabels=true,plainpages=false,colorlinks=true,linkcolor=blue,citecolor=red,urlcolor=blue]{hyperref}
\usepackage[rightcaption]{sidecap}
\usepackage{array}
\usepackage{multirow}
\usepackage{tabularx}
\usepackage{braket} 
\usepackage{dsfont}
\usepackage{bbold}
\usepackage{etoolbox}

\usepackage{dcolumn}
\usepackage{bm}
\usepackage{blindtext}
\usepackage{comment}

\newcommand{\bs}[1]{\boldsymbol{#1}}

\def\Br{{\bs{r}}}

\def\Bk{{\bs{k}}}
\def\BS{{\bs{S}}}
\def\BD{{\bs{D}}}

\def\Bm{{ \bs{m} }}

\def\xhat{{ \bs{\hat{x}} }}
\def\yhat{{ \bs{\hat{y}} }}
\def\zhat{{ \bs{\hat{z}} }}

\def\CalE{{\mathcal{E}}}

\begin{document}


\title{Laser-controlled real- and reciprocal-space topology in multiferroic insulators}

\author{Tomoki Hirosawa}
\affiliation{Department of Physics, University of Tokyo, Bunkyo, Tokyo 113-0033, Japan}
\affiliation{Department of Physics, University of Basel, Klingelbergstrasse 82, CH-4056 Basel, Switzerland}

\author {Jelena Klinovaja}
\affiliation{Department of Physics, University of Basel, Klingelbergstrasse 82, CH-4056 Basel, Switzerland}

\author{Daniel Loss}
\affiliation{Department of Physics, University of Basel, Klingelbergstrasse 82, CH-4056 Basel, Switzerland}

\author{Sebasti{\'a}n A. D{\'i}az}
\affiliation{Department of Physics, University of Basel, Klingelbergstrasse 82, CH-4056 Basel, Switzerland}
\affiliation{Faculty of Physics, University of Duisburg-Essen, 47057 Duisburg, Germany}

\date{\today}

\begin{abstract}
Magnetic materials in which it is possible to control the topology of their magnetic order in real space or the topology of their magnetic excitations in reciprocal space are highly sought-after as platforms for alternative data storage and computing architectures. Here we show that multiferroic insulators, owing to their magneto-electric coupling, offer a natural and advantageous way to address these two different topologies using laser fields. We demonstrate that via a delicate balance between the energy injection from a high-frequency laser and dissipation, single skyrmions---archetypical topological magnetic textures---can be set into motion with a velocity and propagation direction that can be tuned by the laser field amplitude and polarization, respectively. Moreover, we uncover an ultrafast Floquet magnonic topological phase transition in a laser-driven skyrmion crystal and we propose a new diagnostic tool to reveal it using the magnonic thermal Hall conductivity.
\end{abstract}

\maketitle



Historically, manipulating the magnetic order in solids has led to reliable and widely used data storage devices~\cite{tserkovnyakNonlocalMagnetizationDynamics2005}. The recent mainstream embrace of topology as a powerful guiding principle in condensed matter combined with the growing need for alternative platforms for conventional as well as unconventional computing have reinvigorated magnetism research. For instance, magnetic textures whose real-space topology provides them with enhanced stability have garnered attention as potential information carriers in future logic devices and novel, alternative computing architectures~\cite{parkinMagneticDomainWallRacetrack2008, jonietzSpinTransferTorques2010, fertSkyrmionsTrack2013}. Magnons, the quanta of spin waves supported by magnetic materials, can also become topological but in reciprocal space~\cite{vanhoogdalemMagneticTextureinducedThermal2013, shindouTopologicalChiralMagnonic2013, zhangTopologicalMagnonInsulator2013, nakataMagnonicQuantumHall2017, nakataMagnonicTopologicalInsulators2017}. 
Dictated by the bulk-boundary correspondence, bulk topological magnons can result in robust, unidirectional magnonic currents propagating along the sample edges which have been proposed as information conduits for magnonics~\cite{chumakMagnonSpintronics2015, mookChiralHingeMagnons2020}. So far, the study of topological magnetic textures in real space and topological magnons in reciprocal space have developed independently. Therefore, having a single platform capable of supporting both topologies is highly desirable as it can facilitate research on their combined phenomena, and potentially lead to applications that exploit their unique functionalities either simultaneously or selectively.

\begin{figure}[t]
    \centering
    \includegraphics[width=\columnwidth]{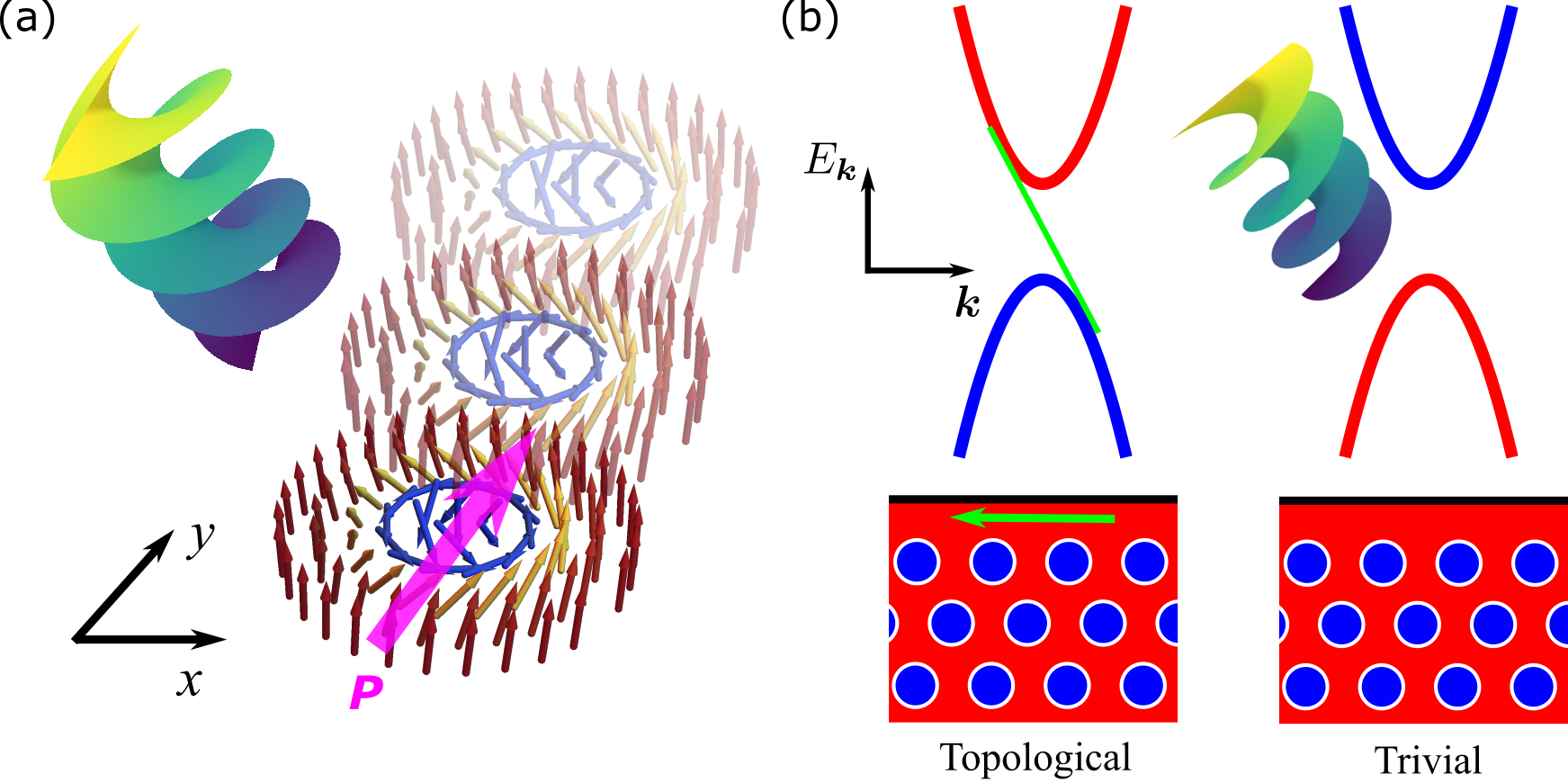}
    \caption{ \textbf{Real- and reciprocal-space topology can be controlled by lasers.} (a) The topological spin structure of a skyrmion carrying in-plane electric polarization $\textbf{P}$ undergoes translational motion under a circularly polarized laser. (b) An ultrafast topological phase transition occurs in the Floquet magnon band structure of skyrmion crystals due to the effective magnetic field induced by circularly polarized laser irradiation.
    }
    \label{fig: schematic}
\end{figure}

Among the family of materials that can support magnetic textures and magnons with nontrivial topology, electrically insulating magnets stand out because they are exempt from Joule heating~\cite{trauzettelAcMagnetizationTransport2008, tuEffectJouleHeating2017}. 
This technologically advantageous property, however, brings about the challenge of controlling magnetism besides the standard approach based on electric currents~\cite{tataraMicroscopicApproachCurrentdriven2008a}. One natural and promising route is to use the electromagnetic field from high-frequency lasers, which can also allow noncontact, ultrafast manipulation~\cite{kirilyukUltrafastOpticalManipulation2010,beaurepaireUltrafastSpinDynamics1996, kimelUltrafastNonthermalControl2005, kampfrathCoherentTerahertzControl2011,takayoshiLaserinducedMagnetizationCurve2014, takayoshiMagnetizationPhaseTransition2014}. Controlling magnetic textures, and hence magnons, with lasers can be accomplished in multiferroics whose general spin lattice Hamiltonian is given by
\begin{equation}
    H(t)=H_0-\sum_{\Br}g\mu_\mathrm{B}\textbf{B}(t)\cdot\BS_{\Br}-\textbf{E}(t)\cdot\textbf{P}_{\Br}\,,
    \label{eq:generalH}
\end{equation}
where $\BS_{\Br}$ is the classical spin vector at site $\Br$, $g$ and $\mu_\mathrm{B}$ respectively denote the g-factor and Bohr magneton, and $H_0$ is the Hamiltonian in the absence of laser irradiation. Here $H(t+T_0) = H(t)$ with $T_0 = 2\pi/\omega_0$ denoting the period of the laser electromagnetic field at frequency  $\omega_0$. Multiferroics not only interact with the magnetic field $\textbf{B}(t)$ from the laser. Thanks to their magnetoelectric coupling their electric dipole moment $\textbf{P}_{\Br}$, induced by the magnetic order, also interacts with the electric field $\textbf{E}(t)$~\cite{dzyaloshinskiiMagnetoElectricalEffectAntiferromagnets1959,kimuraMagneticControlFerroelectric2003a, arimaSpinDrivenFerroelectricityMagnetoElectric2011, satoLaserDrivenMultiferroicsUltrafast2016}. This coupling gives raise to new phenomena and potential functionalities that are not possible just with oscillating magnetic fields.

Here we show that a high-frequency laser, in the terahertz range, can be used to control topological magnetic textures and Floquet topological magnons of multiferroic insulators. A prime example of noncollinear magnetic textures with nontrivial topology in real space are skyrmions~\cite{nagaosaTopologicalPropertiesDynamics2013}. We uncover a novel mechanism that sets skyrmions into motion [see Fig.~\ref{fig: schematic}(a)] where, unlike previous proposals, there is no need to tune the laser frequency to be at resonance with internal skyrmion magnon modes~\cite{wangDrivingMagneticSkyrmions2015, ikkaResonanceModesMicrowavedriven2018, takeuchiSelectiveActivationIsolated2018, yuanWigglingSkyrmionPropagation2019}. The magnon bands supported by a static skyrmion crystal can also be topological in reciprocal space~\cite{roldan-molinaTopologicalSpinWaves2016, garstCollectiveSpinExcitations2017, diazTopologicalMagnonsEdge2019, diazChiralMagnonicEdge2020, hirosawaMagnonicQuadrupoleTopological2020, mookQuantumDampingSkyrmion2020}. We show that a topological phase transition in the Floquet magnon spectrum can be driven by a terahertz laser [see Fig.~\ref{fig: schematic}(b)]. The topological phase reveals itself through robust, chiral magnonic edge states guaranteed by the bulk-boundary correspondence. Furthermore, we establish that the magnonic thermal Hall conductivity~\cite{onoseObservationMagnonHall2010, katsuraTheoryThermalHall2010, matsumotoTheoreticalPredictionRotating2011, vanhoogdalemMagneticTextureinducedThermal2013}, a bulk coefficient, carries a signature of and thus can be used as a diagnostic tool for this topological phase transition. For the above phenomena to occur, we find dissipation to be of crucial importance~\cite{gilbertPhenomenologicalTheoryDamping2004}. To consistently incorporate dissipative effects into our description, thus circumventing the issue of reaching a thermal state at long times, we adopt the Floquet-Magnus expansion for classical systems~\cite{higashikawaFloquetEngineeringClassical2018}, and extend the Floquet magnon formalism to multiferroics with a time-dependent, noncollinear magnetic unit cell.



{\it Model.---}Inspired by the vast body of available experimental characterization, we use a model that closely describes the multiferroic insulator Cu$_2$OSeO$_3$~\cite{sekiObservationSkyrmionsMultiferroic2012, sekiMagnetoelectricNatureSkyrmions2012,mochizukiDynamicalMagnetoelectricPhenomena2015}, which is known to host skyrmions as isolated objects as well as in crystalline form, hence an ideal testbed for our predictions.

Considering a thin film sample, the spin lattice Hamiltonian is defined on a square lattice as~\cite{mochizukiSpinWaveModesTheir2012, mochizukiMagnetoelectricResonancesPredicted2013, mochizukiMicrowaveMagnetochiralEffect2015}
\begin{eqnarray}
\label{eq: SpinLatticeH_floquet}
H_0&=&\frac{1}{2}\sum_{\braket{\Br,\Br'}}(-J_{\Br,\Br'}\BS_{\Br}\cdot \BS_{\Br'}+ \BD_{\Br,\Br'}\cdot\BS_{\Br}\times\BS_{\Br'})\nonumber\\
&-&g\mu_{B} B_0 \sum_{\Br} \BS_{\Br}\cdot\zhat \,.
\end{eqnarray}
The nearest-neighbor spins interact via ferromagnetic exchange $J_{\Br,\Br'} = J(\delta_{\Br-\Br',\pm a\xhat}+\delta_{\Br-\Br',\pm a\yhat})$, with $J>0$ and $a$ is the lattice constant, and Dzyaloshinskii-Moriya interaction $\boldsymbol{D}_{{\boldsymbol{r},\boldsymbol{r}'}}=D(\boldsymbol{r}-\boldsymbol{r}')/|\boldsymbol{r}-\boldsymbol{r}'|$. In what follows, $\BS_{\Br}=S\Bm_{\Br}$ is the total magnetization at site $\Br$ with $\Bm_{\Br}$ denoting a unit vector and $S=1$ for each Cu-ion tetrahedra~\cite{mochizukiMagnetoelectricResonancesPredicted2013}. The spins also couple to an applied static magnetic field $\textbf{B}_0 = B_0\zhat$.

In Cu$_2$OSeO$_3$, a noncollinear spin texture induces a local electric polarization via the $d$-$p$ hybridization mechanism~\cite{arimaSpinDrivenFerroelectricityMagnetoElectric2011, sekiMagnetoelectricNatureSkyrmions2012, mochizukiMagnetoelectricResonancesPredicted2013} and whose explicit form depends on the direction of the applied static magnetic field relative to the crystallographic axes. Orienting the sample such that $\textbf{B}_0\parallel [110]$, the local electric dipole moment is given by~\cite{sekiMagnetoelectricNatureSkyrmions2012, mochizukiMagnetoelectricResonancesPredicted2013}
\begin{equation}
    \textbf{P}_{\Br}=\lambda(- m_{\Br,x}m_{\Br,y},\frac{-m_{\Br,x}^2+m_{\Br,z}^2}{2},m_{\Br,y}m_{\Br,z}),
    \label{eq:P[110]}
\end{equation}
thus inducing a total electric polarization along the $y$-axis and where the magnetoelectric coupling strength is $\lambda=5.64\times 10^{-27}\mu$Cm.

The time-periodic driving fields in Eq.~\eqref{eq:generalH} are introduced as $\textbf{E}(t)=E_d\{\sin(\omega_0 t+\delta),\cos(\omega_0 t),0\}$ and $\textbf{B}(t)=B_d\{\cos(\omega_0 t),-\sin(\omega_0 t+\delta),0\}$ with $B_d=E_d/c$ and $c$ denoting the speed of light.
The laser polarization is determined by $\delta$, with $\delta=0,\pi/2,\pi$ corresponding to right-circularly polarized (RCP), linearly polarized, and left-circularly polarized (LCP) light, respectively. In Table~\ref{table: unit_conversion}, we summarize the conversion between dimensionless parameters and physical units for this model.

\begin{table}[t]
  \centering
  \begin{tabular}{c|c|c}
  &Dimensionless&Physical Unit\\ \hline 
  \rule{0pt}{4ex}    Time & $\overline{t}=\hbar t/JS$ & $\approx$ 0.66 ps\\
  Distance & $\overline{x}=x/a$ & 0.5 nm\\  
  Magnetic field (static)& $\overline{B}_{0}=g\mu_{\mathrm{B}}B_0J/D^2 $ & $\approx$ 0.07 T\\
  Magnetic field (laser)& $\tilde{B}_d=g\mu_{\mathrm{B}}B_d/D $ & $\approx$ 0.78 T\\
  Frequency (spin waves) & $\overline{\omega}=\hbar \omega J/ D^2$ & $\approx $ 2.0 GHz \\
  Frequency (laser) & $\tilde{\omega}_0=\hbar \omega_0 /J$ & $\approx $ 250 GHz 
 \end{tabular}
 \caption{Unit conversion table for $J=1\textrm{ meV}$, $D/J=0.09$ and $a=0.5\textrm{ nm}$~\cite{mochizukiDynamicalMagnetoelectricPhenomena2015}. Physical variables are obtained by multiplying the dimensionless variables on the middle column by the values on the right column.}
 \label{table: unit_conversion}
\end{table}

To describe the nonequilibrium steady state of the periodically driven magnetization, we employ the high frequency Floquet-Magnus expansion~\cite{higashikawaFloquetEngineeringClassical2018} on the Landau-Lifshitz-Gilbert (LLG) equation and
obtain the following effective Floquet Hamiltonian
\begin{eqnarray}
H_F&=&H_0+\sum_{\Br} g\mu_{\mathrm{B}}B_F\Big[-\hat{z}\cdot \Bm_{\Br}+\mathcal{E}_d m_{\Br,z}m_{\Br,x}\nonumber\\
&-&\mathcal{E}_d^2(-2m_{\Br,x}^2+m_{\Br,y}^2)\hat{z}\cdot \Bm_{\Br}\Big],\quad\quad
\label{eq:Floquet_SpinLatticeH}
\end{eqnarray}
where $B_{F}=\frac{\gamma B_d^2\cos\delta }{2(1+\alpha^2)\omega_0}$ with $\gamma$ the gyromagnetic ratio and $\alpha$ the Gilbert damping constant, and $\mathcal{E}_d=  \lambda c / g\mu_{B}$~(Supplementary Material \cite{SM}).
The second term corresponds to the effective magnetic field~\cite{takayoshiLaserinducedMagnetizationCurve2014, takayoshiMagnetizationPhaseTransition2014, higashikawaFloquetEngineeringClassical2018, miyakeCreationNanometricMagnetic2020a}, while the third and fourth terms arise from the magnetoelectric nature of Cu$_2$OSeO$_3$. Below we show how these terms give rise 
to a novel control of the topology in real and reciprocal space.

\begin{figure}[t]
    \centering
    \includegraphics[width=\columnwidth]{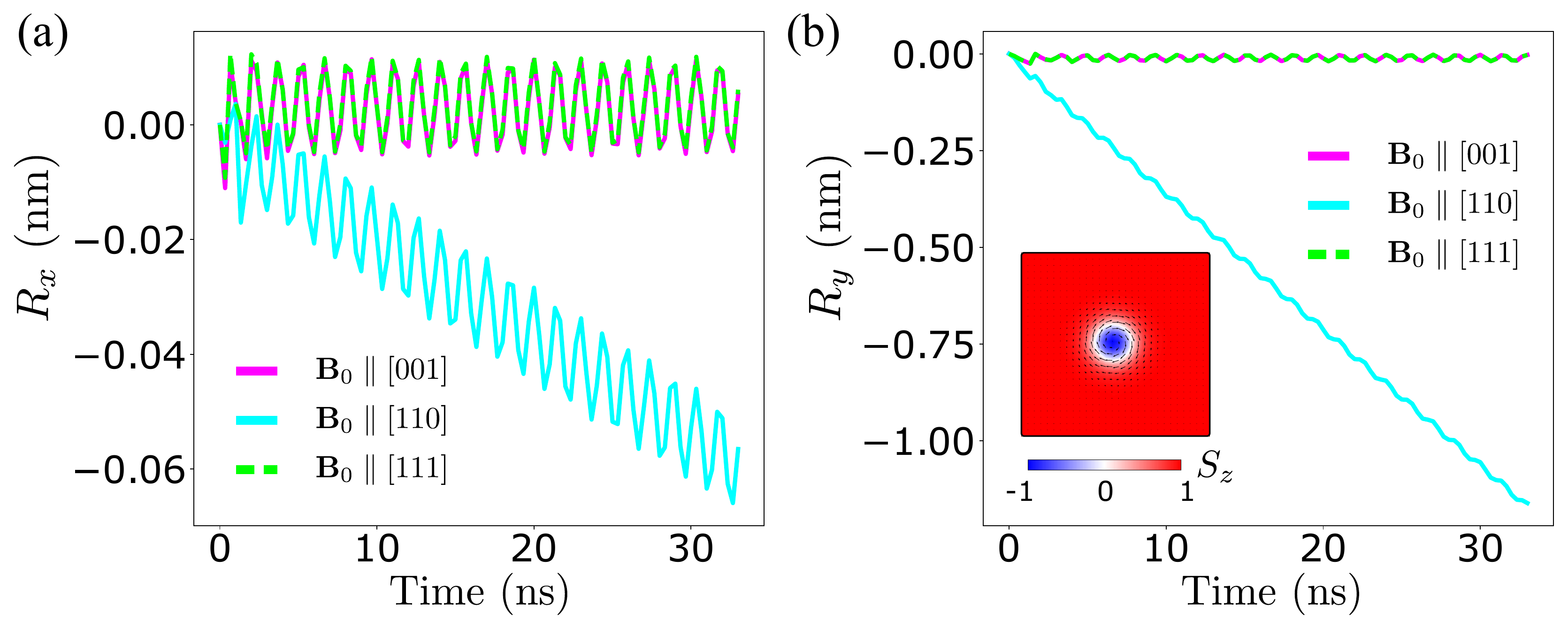}
    \caption{ \textbf{Skyrmion motion under a circularly polarized laser.} (a), (b) Displacement of the center of skyrmions (a) $R_x$ and (b) $R_y$ under LCP light as a function of time, defined in the text. The result for $\textbf{B}_0\parallel [001]$, $\textbf{B}_0\parallel [110]$, and $\textbf{B}_0\parallel [111]$ is respectively depicted in solid magenta, solid cyan, and dashed lime lines. The nonlinear effect of damping and asymmetric magnetoelectric coupling drive skyrmions without exciting internal magnon modes.
    The parameters are fixed as $D/J=0.09$, $\overline{B}_0=0.9$, $\tilde{B}_d=2.0$, $\tilde{\omega}_0=10$, $\CalE_d=0.1$, and $\alpha=0.04$ (see Table~\ref{table: unit_conversion}).
    The inset in (b) shows the initial configuration of a single skyrmion prepared by Monte Carlo annealing for $150\times 150$ spins~\cite{evansAtomisticSpinModel2014}.
     }
    \label{fig: motion_axis}
\end{figure}



{\it Laser-driven skyrmion motion.---}Since the $x$-component of the magnetization changes sign across the center of skyrmions, the third term of Eq.~\eqref{eq:Floquet_SpinLatticeH}, obtained for $\textbf{B}_0\parallel [110]$, induces a dipole-like distortion in the spin texture~\cite{SM}. Crucially, the induced distortion breaks the skyrmions rotational symmetry, thus allowing for their laser-driven motion. 

First we consider the dynamics of a single skyrmion under LCP light. We numerically solve the LLG equation obtained from the time-dependent Hamiltonian in Eq.~\eqref{eq:generalH}~\cite{SM}.
The motion of a single skyrmion is tracked by computing its center, defined as $R_{i}=\int d\Br^2 \rho(\Br) r_i /Q $ for $i=x,y$ with $\rho(\Br)$ and $Q=\int d\Br^2 \rho(\Br)$ denoting the topological charge density and total topological charge, respectively~\cite{bergDefinitionStatisticalDistributions1981}. 

The change in $R_x$ and $R_y$ as functions of time after irradiation of LCP light is shown in Fig.~\ref{fig: motion_axis}. Orienting $\textbf{B}_0$ along other crystallographic directions induces different local electric dipole moments ~\cite{SM, sekiMagnetoelectricNatureSkyrmions2012, mochizukiDynamicalMagnetoelectricPhenomena2015}. 
We find that the skyrmion under $\textbf{B}_0\parallel [110]$ exhibits translational motion along the negative $y$-axis and a small drift velocity along the negative $x$-axis, hence the skyrmion motion is mostly directed antiparallel to the net electric polarization.
In contrast, the results for $\textbf{B}_0\parallel [001], [111]$ show no net displacement with almost identical trajectories. 
This is consistent with the symmetry argument, since the time-averaged spin textures for $\textbf{B}_0\parallel [001]$ and $\textbf{B}_0\parallel[111]$ respectively display $C_{2z}$ and $C_{3z}$ rotational symmetry, while no in-plane rotational symmetry is present for $\textbf{B}_0\parallel [110]$~\cite{SM}.

We then combine the Floquet-Magnus expansion~\cite{higashikawaFloquetEngineeringClassical2018} with Thiele's approach to derive the analytical expression of the skyrmions drift velocity $\boldsymbol{v}$~\cite{thieleSteadyStateMotionMagnetic1973, wangDrivingMagneticSkyrmions2015}.
Up to first order in the Floquet-Magnus expansion, we obtain
\begin{align}
v_x&\approx -\frac{\alpha\eta}{4\pi Q} v_y,\nonumber\\
v_y&\approx -\frac{2.25}{Q} v_0\overline{B}_F\CalE_d\alpha ,\label{eq: analytical_Velocity}
\end{align}
where $\eta\approx 4\pi$ depends on the skyrmions spatial profile, $v_0=a\gamma  D^2/g\mu_BJ\approx 6.16$ m/s is a characteristic velocity, and $\overline{B}_F=g\mu_B B_FJ/D^2$ is the dimensionless effective magnetic field.
It is important to note that the drift velocity is in general proportional to the damping constant for an arbitrary local dipole moment $P_{\Br,i}=\chi^{i}_{jk}m_{\Br,j}m_{\Br,k}$ with $i,j,k=x,y,z$~\cite{SM}. Hence, damping is essential for the off-resonant skyrmion motion, where no internal magnon modes are excited due to the large difference between the laser THz-range frequency and the skyrmion magnon eigenfrequencies.

From Fig.~\ref{fig: motion_axis}, the velocity for $\textbf{B}_0\parallel[110]$ is estimated as $(v_x,v_y)=(-0.17\textrm{~cm/s},-3.52\textrm{~cm/s})$.
In comparison, Eq.~\eqref{eq: analytical_Velocity} gives $(v_x,v_y)=(-0.088\textrm{~cm/s}, -2.22\textrm{~cm/s})$ for $Q=-1$ and $\eta=4\pi$.
On the other hand, the equivalent calculation for $\textbf{B}_0\parallel[001],[111]$ yields $\boldsymbol{v}=0$ for the leading order term proportional to $\omega_0^{-1}$~\cite{SM}.
Hence, our analytical expression is consistent with the LLG simulations. 
The dependence of $v_y$ for $\textbf{B}_0\parallel[110]$ on $B_d, \mathcal{E}_d, \omega_0, $ and $\alpha $ is discussed in detail in~\cite{SM}, which is shown to be in a good agreement with Eq.~\eqref{eq: analytical_Velocity} for small $B_d$ and large $\omega_0$, where the asymptotic behavior is described by the high frequency expansion. Furthermore, our simulations confirm the skyrmion motion under ultrashort pulses~\cite{SM} and reversal of skyrmion direction with switching from LCP light to RCP light, which changes the sign of $B_F$.
We have demonstrated that the amplitude, frequency, and chirality of the laser affect the off-resonant motion of skyrmions, thus providing a greater control on their velocity, which can achieve a maximum of 15 cm/s~\cite{SM}.



{\it Floquet magnonic topological phase transition.---}On top of the applied static field $B_0$, circularly polarized laser irradiation generates an effective magnetic field $B_F$, which can be tuned by the laser field amplitude $B_d$. Therefore, analogously to the static magnetic-field driven topological phase transition in skyrmion crystals~\cite{diazChiralMagnonicEdge2020}, we show that a Floquet magnonic topological phase transition can be driven by a circularly-polarized laser.

Using the LLG equation, we study the magnetic excitations of a non-equilibrium steady-state skyrmion crystal sustained by RCP laser irradiation. We introduce, in addition to the laser field, pulsed magnetic fields along the in-plane and out-of-plane directions, which excite magnetically active skyrmion crystal spin wave modes~\cite{mochizukiSpinWaveModesTheir2012}. The resonance frequencies of these modes are obtained from the dynamical susceptibilities~\cite{mochizukiSpinWaveModesTheir2012, mochizukiMicrowaveMagnetochiralEffect2015} $\textrm{Im}\chi_{ij}(\omega)=M_{\omega,i}/B_{\omega,j}$, with $M_{\omega,i}$ and $B_{\omega,j}$ denoting the Fourier transform of the total magnetization $\boldsymbol{M}(t)$ and $\textbf{B}(t)$ for $i,j=x,y,z$~\cite{SM}.

\begin{figure}[t]
    \centering
    \includegraphics[width=\columnwidth]{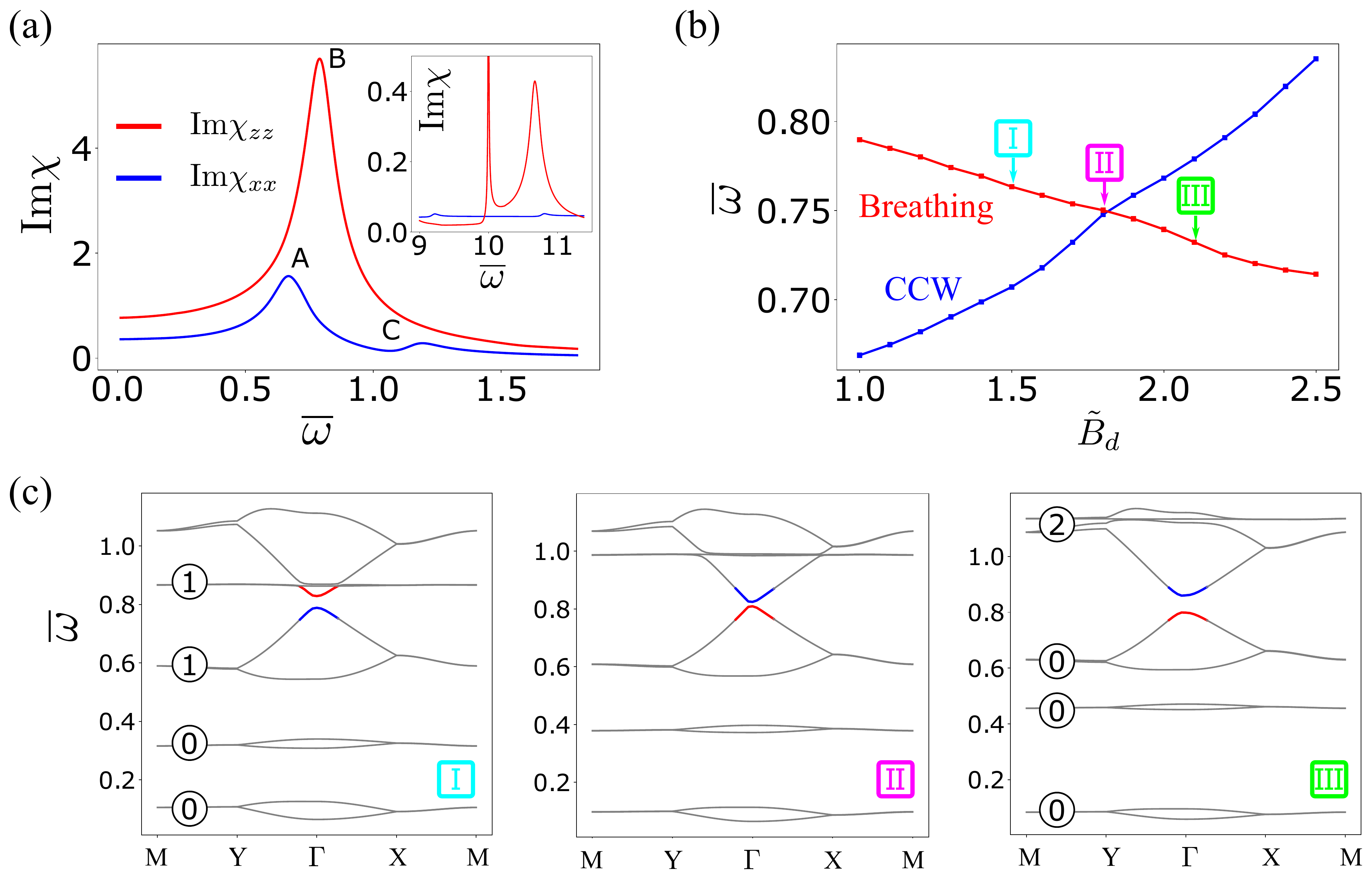}
    \caption{
\textbf{Floquet magnonic topological phase transition in skyrmion crystals.} (a) Imaginary parts of in-plane (out-of-plane) dynamical susceptibilities shown in blue (red), obtained at $D/J=1.0$, $\overline{B}_0=0.4$, $\tilde{B}_d=1.0$, $\tilde{\omega}_0=10$, $\mathcal{E}_d=0.1$, and $\alpha=0.04$ under RCP light. The resonance peaks of the (A) counterclockwise (CCW), (B) breathing, and (C) clockwise (CW) modes are indicated. The inset shows the result near the driving frequency $\tilde{\omega}_0=10$. (b) Resonance frequencies of the CCW (blue) and breathing (red) modes as functions of the laser field amplitude $\tilde{B}_d$. (c) Floquet magnon band structures in the vicinity of the topological phase transition with integers indicating the total Chern number for each group of bands. The CCW and breathing modes at the $\Gamma$ points are highlighted in blue and red, respectively. Roman numerals in (b) mark the values of $\tilde{B}_d$ used in the panels in (c).
     }
    \label{fig: floq_magnon}
\end{figure}

Figure~\ref{fig: floq_magnon}(a) shows $\textrm{Im}\chi_{xx}$ (blue) and $\textrm{Im}\chi_{zz}$ (red), where the resonance peaks of the counterclockwise (CCW), breathing, and clockwise (CW) modes are labeled as A, B, and C, respectively. These low-energy modes are similar to those obtained in static systems~\cite{mochizukiSpinWaveModesTheir2012}. Additionally, a mode is found at $\overline{\omega}\approx 10.8$ in $\textrm{Im}\chi_{zz}$ as illustrated in the inset of Fig.~\ref{fig: floq_magnon}(a). This high-energy mode is expected from the periodic structure of the quasienergy spectrum given by $\epsilon_{nm}=\epsilon_n+m\hbar\omega_0$ for eigenvalues $\epsilon_n$ of the effective Floquet Hamiltonian and integers $n,m$~\cite{eckardtHighfrequencyApproximationPeriodically2015}. We should note that the large, sharp peak at $\omega=\omega_0$ in $\textrm{Im}\chi_{zz}$ corresponds to the laser field, while the signal from $\textrm{Im}\chi_{xx}$ is too weak to observe any peaks near $\omega=\omega_0$.

The resonance frequencies of the CCW (blue) and breathing (red) modes as functions of $\tilde{B}_d$ are shown in Fig.~\ref{fig: floq_magnon}(b), from where we identify the band inversion point at the critical value $\tilde{B}_d\approx 1.8$.
The total effective magnetic field at this point is estimated as $\overline{B}_F+\overline{B}_0\approx 0.56$.
Remarkably, it is almost equal to the critical magnetic field obtained in static systems~\cite{diazChiralMagnonicEdge2020} and thus consistent with Floquet magnons experiencing a laser-induced effective field. 
We find that the critical value of the total effective magnetic field remains roughly constant up to $\tilde{\omega}_0\ge 5$, where the high frequency expansion is justified~\cite{SM}.

Treating the time-averaged spin configuration of the non-equilibrium steady state as the classical spin ground-state for the effective Floquet Hamiltonian in Eq.~\eqref{eq:Floquet_SpinLatticeH}, we compute the Floquet magnon band structures (see~\cite{SM}) shown in Fig.~\ref{fig: floq_magnon}(c). The laser-driven Floquet magnonic topological phase transition is signaled by the closing of the gap between the CCW and breathing modes, at the band inversion point, with the consequent Chern number transfer.  
The obtained Floquet magnon band structures are consistent with the LLG simulation results.
We note that the topological phase transition occurs at a slightly smaller $\tilde{B}_d$ value in the Floquet magnon band structures, which may arise from higher order terms neglected in the Floquet Hamiltonian. Invoking the bulk-boundary correspondence, the above Chern number transfer opens the door to laser-controlled switching of nonreciprocal Floquet magnonic chiral edge states due to the magnetoelectric coupling~\cite{SM}.

\begin{figure}[t]
    \centering
    \includegraphics[width=60mm]{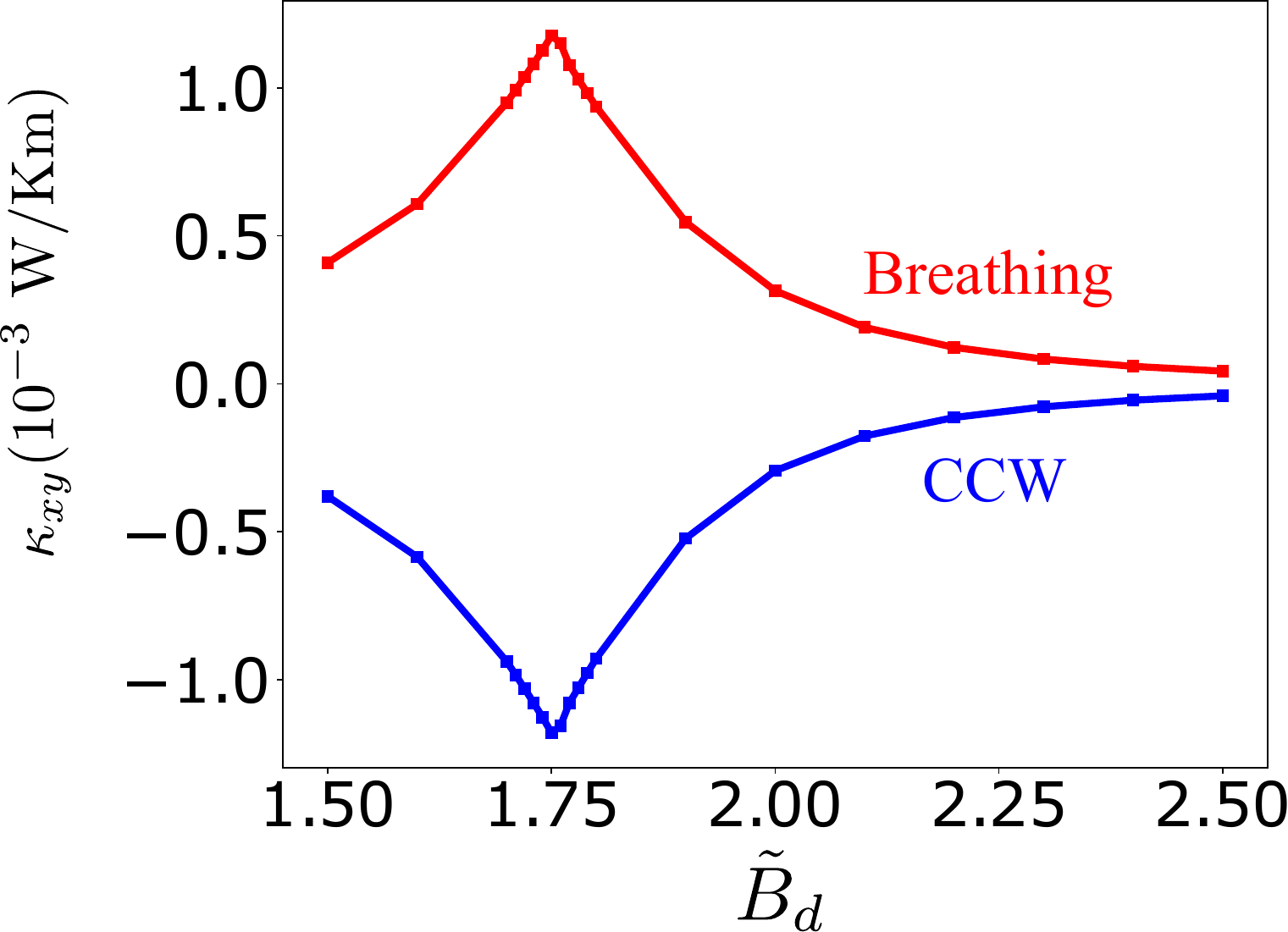}
    \caption{
\textbf{Diagnosing a Floquet magnonic topological phase transition via the magnonic thermal Hall effect.} 
The magnonic thermal Hall conductivity of the CCW (blue) and breathing (red) modes plotted against the laser field amplitude. These two modes can be resonantly and selectively excited using ac magnetic fields. The opposite sign enhancement at the critical value, due to the divergence of the Berry curvature, can be used as a topological phase transition diagnostic tool. The parameters are the same as in Fig~\ref{fig: floq_magnon}(b).
     }
    \label{fig: thermal_Hall}
\end{figure}


As a bulk probe for the laser-driven topological phase transition, we consider the Floquet magnonic thermal Hall conductivity~\cite{katsuraTheoryThermalHall2010, onoseObservationMagnonHall2010, matsumotoTheoreticalPredictionRotating2011, owerreFloquetTopologicalMagnons2017}
\begin{equation}
\kappa_{xy}=-\frac{k_B^2T}{\hbar d}\sum_{n}\int \frac{d^2\boldsymbol{k}}{(2\pi)^2} c_2(\rho_{n,\boldsymbol{k}})\Omega_{n}(\boldsymbol{k}),
\label{eq: thermal_Hall}
\end{equation}
where $d$ is the interlayer distance, $\Omega_{n}(\boldsymbol{k})$ is the Berry curvature of the $n$-th Floquet magnon band, $\rho_{n,\boldsymbol{k}}$ is the distribution function of the $n$-th Floquet magnon eigenstate at $\Bk$, and $c_{2}(\rho)=(1+\rho)(\log\frac{1+\rho}{\rho})^2-(\log\rho)^2-2\textrm{Li}_2(-\rho)$ with $\textrm{Li}_2(z)$ denoting the polylogarithm function of order two. 
While the population of magnons in the CCW/breathing mode is small at thermal equilibrium, we could selectively pump them with in-plane/out-of-plane ac magnetic fields~\cite{mochizukiSpinWaveModesTheir2012, onoseObservationMagneticExcitations2012,garstCollectiveSpinExcitations2017}.
In this resonance condition, we can approximate $\rho_{n,\boldsymbol{k}}$ with a Gaussian distribution sharply peaked at the $\Gamma$ point of the corresponding magnon band~\cite{SM}.
Crucially, the thermal Hall conductivity from the resonant CCW/breathing mode is strongly enhanced due to the divergence of the Berry curvature at the topological phase transition. 
Figure~\ref{fig: thermal_Hall} clearly illustrates the enhancement with opposite sign for the CCW and breathing modes, obtained for a thin film sample of Cu$_2$OSeO$_3$ with $d=1$~nm and $T=5$~K~\cite{sekiObservationSkyrmionsMultiferroic2012}.

{\it Conclusion.---}We have established a novel formalism to describe laser-controlled topology in real- and reciprocal-space of multiferroic insulators. Crucially, our theory incorporates the phenomenological damping effect to the Floquet magnon formalism, which is essential in a non-equilibrium condition under driving fields. Taking Cu$_2$OSeO$_3$ as a model, we have demonstrated the off-resonance motion of skyrmions due to the magnetoelectric effect. We have also shown that circularly polarized lasers induce a Floquet magnonic topological phase transition. Furthermore, exploiting the selection rules of CCW and breathing modes of skyrmion crystals~\cite{onoseObservationMagneticExcitations2012,mochizukiSpinWaveModesTheir2012,garstCollectiveSpinExcitations2017}, we have suggested the magnonic thermal Hall conductivity as a diagnostic tool for the topological phase transition. Our formalism can be generalized to other multiferroics such as GaV$_4$S$_8$~\cite{kezsmarkiNeeltypeSkyrmionLattice2015, ruffMultiferroicitySkyrmionsCarrying2015}. The ultrafast control of topological spin structures and Floquet magnon band topology in multiferroic insulators broadens the spectrum of potential spintronic and magnonic applications.

%
%


\begin{acknowledgments}
We are grateful to T. Hinokihara, H. Matsuura, and M. Ogata for discussions. We also thank S. Seki and A. Mook for their insightful advice on the magnonic thermal Hall effect in skyrmion crystals. 
T. H. is supported by Japan Society for the Promotion of Science (JSPS) through Program for Leading Graduate Schools (MERIT) and JSPS KAKENHI (Grant No. 18J21985). This work was supported by the Swiss National Science Foundation and NCCR QSIT. This project received funding from the European Union's Horizon 2020 research and innovation program (ERC Starting Grant, Grant Agreement No. 757725).
\end{acknowledgments}





%

\end{document}


\title{Supplemental Material for ``Laser-controlled real and reciprocal space topology in multiferroic insulators''}

\author{Tomoki Hirosawa}
\affiliation{Department of Physics, University of Tokyo, Bunkyo, Tokyo 113-0033, Japan}
\affiliation{Department of Physics, University of Basel, Klingelbergstrasse 82, CH-4056 Basel, Switzerland}

\author {Jelena Klinovaja}
\affiliation{Department of Physics, University of Basel, Klingelbergstrasse 82, CH-4056 Basel, Switzerland}

\author{Daniel Loss}
\affiliation{Department of Physics, University of Basel, Klingelbergstrasse 82, CH-4056 Basel, Switzerland}

\author{Sebasti{\'a}n A. D{\'i}az}
\affiliation{Department of Physics, University of Basel, Klingelbergstrasse 82, CH-4056 Basel, Switzerland}
\affiliation{Faculty of Physics, University of Duisburg-Essen, 47057 Duisburg, Germany}


\maketitle



\section{Numerical simulation of the LLG equation}

Under laser irradiation, the classical spin dynamics of multiferroic insulators is described by the Landau-Lifshiz-Gilbert (LLG) equation at zero temperature:
\begin{equation}
    \frac{d\Bm_{\Br}}{dt}=-\frac{\gamma \Bm_{\Br}}{1+\alpha^2}\times\left[ \BH_{\Br}^\textrm{eff}(t) + \alpha\Bm_{\Br}\times\BH_{\Br}^\textrm{eff}(t) \right] \,, 
    \label{eq: general_LLG}
\end{equation}
where $\Bm_\Br=\BS_{\Br}/S$ is the normalized magnetization, $\BH_{\Br}^\textrm{eff}(t) = -[1/(\hbar \gamma S)]\partial H(t)/\partial \Bm_{\Br}$ with $H(t+T_0)=H(t)$, $\gamma = g\mu_{\rm{B}}/ \hbar$ is the gyromagnetic ratio, and $\alpha$ is the Gilbert damping constant.
We fix the damping constant $\alpha=0.04$ and the laser frequency $\tilde{\omega}_0=10$ unless otherwise stated.
As the initial condition for the LLG equation we take the ground-state spin texture of the static Hamiltonian $H_0$ defined in the main text, which is obtained by Monte Carlo annealing under periodic boundary conditions~\cite{evansAtomisticSpinModel2014}. 
The LLG equation is solved numerically with a discretized time step $\Delta \overline{t}=10^{-2}$. 
After a sufficiently long relaxation time, we obtain the non-equilibrium steady state characterized by time-periodic oscillations of the total magnetization.


\section{Electric polarization of C\titlelowercase{u}$_2$OS\titlelowercase{e}O$_3$ }
\label{section: multiferroics}
Following~Refs.~\cite{sekiMagnetoelectricNatureSkyrmions2012, mochizukiDynamicalMagnetoelectricPhenomena2015}, we derive expressions for the local electric polarization in Cu$_2$OSeO$_3$ for three different directions of the applied magnetic field, namely, $\textbf{B}_0\parallel[001],[110]\textrm{, and }[111]$. 
For this material, it was shown that the $d-p$ hybridization mechanism between magnetic ions and ligand anions provides the correct relationship between the experimentally measured electric polarization and the direction of the applied magnetic field~\cite{arimaFerroelectricityInducedProperScrew2007, sekiMagnetoelectricNatureSkyrmions2012}.
The expression of local electric polarization was obtained as~\cite{sekiMagnetoelectricNatureSkyrmions2012}
\begin{eqnarray}
\textbf{P}_{\Br,abc}(\Bm_{\Br,abc})&=&(p_{\Br,a},p_{\Br,b},p_{\Br,c})\nonumber\\
&=&\lambda(m_{\Br,b}m_{\Br,c},m_{\Br,c}m_{\Br,a},m_{\Br,a}m_{\Br,b}),\quad\quad
\end{eqnarray}
with $\lambda$ denoting the magnetoelectric coupling constant given in the main text and the indices $a,b,c$ indicate that it is defined in the orthogonal basis of cubic crystalline axes with $\hat{a}\parallel[100]$, $\hat{b}\parallel[010]$, and $\hat{c}\parallel[001]$.

For convenience, we introduce another orthogonal basis $(x,y,z)$, which is used throughout the main text, with the $z$-axis parallel to the applied magnetic field and the $x$-axis is taken along the [$\overline{1}$10] direction.
Now, we define the rotation matrix $\hat{R}_{abc,xyz}$ which maps from $xyz$-coordinates to $abc$-coordinates
\begin{equation}
\Bm_{\Br,abc}=\hat{R}_{abc,xyz}\Bm_{\Br,xyz},
\end{equation}
where $\Bm_{\Br,abc}$/$\Bm_{\Br,xyz}$ is the normalized magnetization defined in $abc$-/$xyz$-coordinates.
The local electric polarization in $xyz$-coordinates is written as
\begin{equation}
\textbf{P}_{\Br,xyz}(\Bm_{\Br,xyz})=\hat{R}_{abc,xyz}^{-1}\textbf{P}_{\Br,abc}(\hat{R}_{abc,xyz}\Bm_{\Br,xyz}).
\label{eq: rotated_polarization}
\end{equation}

Using Eq.~\eqref{eq: rotated_polarization}, the following local electric polarizations in $xyz$-coordinates are obtained
\begin{align}
\textbf{P}_{\Br,xyz}&=\lambda\Big(- m_{\Br,z}m_{\Br,x},m_{\Br,y}m_{\Br,z},\frac{-m_{\Br,x}^2+m_{\Br,y}^2}{2}\Big)\nonumber\\
&\quad\quad\quad\quad\textrm{for }\textbf{B}_0\parallel [001],\label{eq: polarization [001]}\\
\textbf{P}_{\Br,xyz}&=\lambda\Big(- m_{\Br,x}m_{\Br,y},\frac{-m_{\Br,x}^2+m_{\Br,z}^2}{2},m_{\Br,y}m_{\Br,z}\Big)\nonumber\\
&\quad\quad\quad\quad\textrm{for }\textbf{B}_0\parallel [110],\label{eq: polarization [110]}\\
\textbf{P}_{\Br,xyz}&=\lambda\Big(-\frac{m_{\Br,x}(\sqrt{2}m_{\Br,y}+m_{\Br,z})}{\sqrt{3}}, \nonumber\\
&\frac{-m_{\Br,x}^2+m_{\Br,y}(m_{\Br,y}-\sqrt{2}m_{\Br,z})}{\sqrt{6}},\nonumber\\ &-\frac{m_{\Br,x}^2+m_{\Br,y}^2-2m_{\Br,z}^2}{2\sqrt{3}}\Big)\textrm{ for }\textbf{B}_0\parallel [111].\label{eq: polarization_all_basis}
\end{align}
Assuming $\Bm_{\Br,xyz}=(0,0,1)$ for the ferromagnetic phase, we immediately obtain $\textbf{P}_{\Br,xyz}=0$ for $\textbf{B}_0\parallel [001]$, $\textbf{P}_{\Br,xyz}=\frac{\lambda}{2}\hat{y}$ for $\textbf{B}_0\parallel [110]$ (with $\hat{y}\parallel $[001]), and $\textbf{P}_{\Br,xyz}=\frac{\lambda}{\sqrt{3}}\hat{z}$ for $\textbf{B}_0\parallel [111]$. In what follows, we exclusively work with $xyz$-coordinates.

Let $\hat{U}_S$ represent a symmetry operation of the skyrmion spin configuration. The local electric polarization respects this symmetry when 
\begin{equation}
    \hat{U}_SP_{\Br}(\Bm_\Br)=P_{\Br}(\hat{U}_S\Bm_{\Br}).
\end{equation}
However, this is not necessarily the case. In fact, only a subgroup of the rotational symmetries of skyrmions is respected by the local electric polarization. 
For example, the twofold rotational symmetry about the $z$-axis, $C_{2z}$, transforms magnetic dipole moments as
\begin{equation}
    (m_{x},m_y,m_z)_{[x,y,z]}\overset{C_{2z}}{\rightarrow} (-m_x,-m_y,m_z)_{[-x,-y,z]},
\end{equation} 
with the square bracketed indices denoting the position vector.
Given a $C_{2z}$ symmetric spin configuration, the symmetry is preserved in the spatial profile of local electric polarization only if
\begin{align}
    (P_x,P_y,P_z)_{[x,y,z]}&\overset{C_{2z}}{\rightarrow}(-P_x,-P_y,P_z)_{[-x,-y,z]}\nonumber\\
    &=P_{\Br}\Big((-m_x,-m_y,m_z)_{[-x,-y,z]}\Big).
\end{align}
From Eq.~\eqref{eq: polarization [001]}-\eqref{eq: polarization_all_basis}, this condition is satisfied only for $\textbf{B}_0\parallel [001]$. Similarly, it is straightforward to show that the $C_{3z}$ symmetry is respected only for $\textbf{B}_0\parallel [111]$. In contrast, rotational symmetry is completely lost for $\textbf{B}_0\parallel [110]$. 

\begin{figure}[t]
    \centering
    \includegraphics[width=\columnwidth]{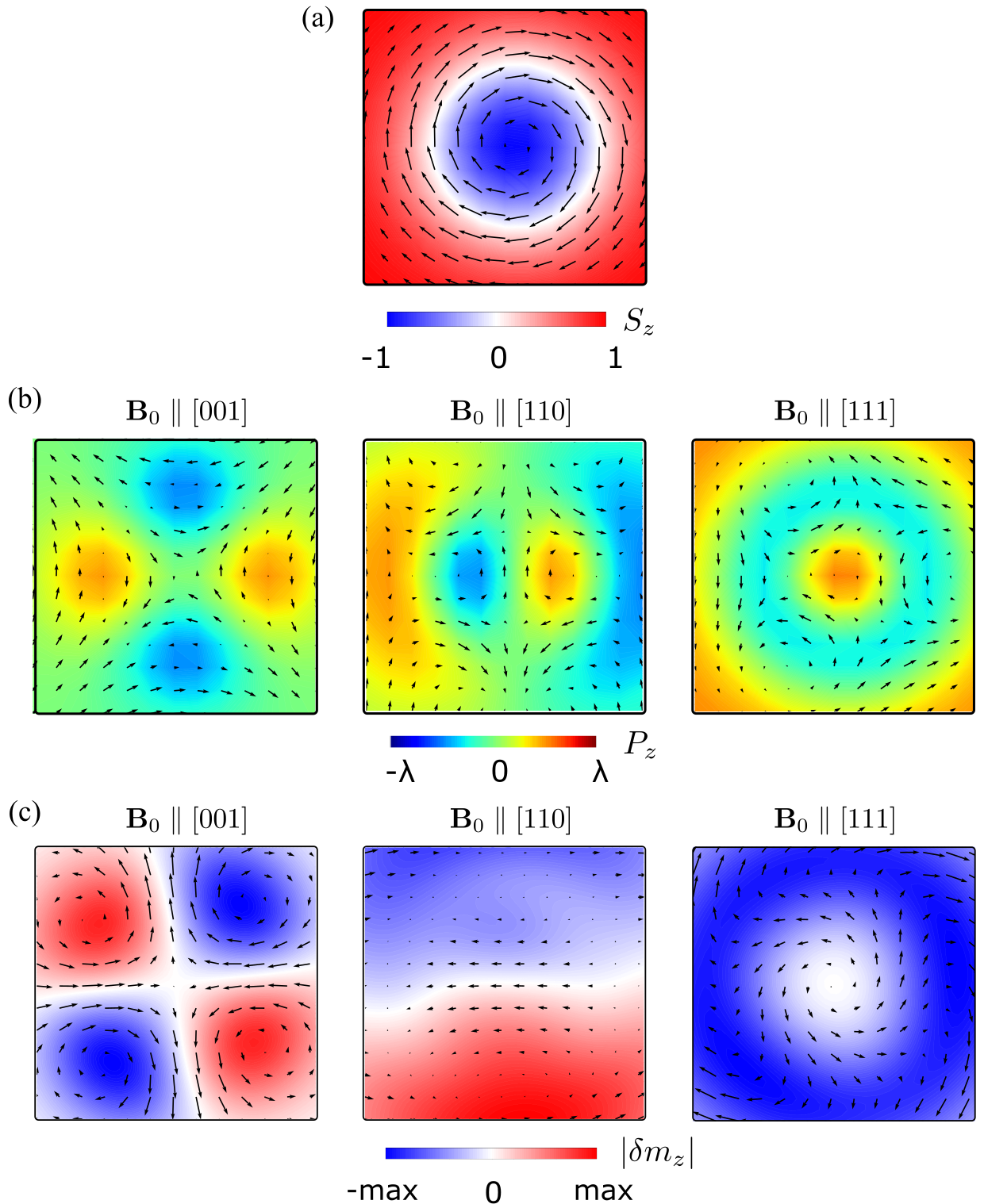}
    \caption{ \textbf{Coupling between local electric polarization and laser induces the distortion of spin textures. }
(a)~Enlarged spin configuration of a single skyrmion, obtained at $D/J=0.09$ and $\overline{B}_{0}=0.9$ for 150$\times$150 spins with periodic boundary conditions. (b) Spatial profile of local electric polarization $\textbf{P}_{\Br}$ inside the skyrmion. (c) Induced distortion $\delta\Bm_{\Br}$ of the skyrmion under the left-circularly polarized laser (LCP), obtained by the effective LLG equations at $\tilde{B}_d=2.0$, $\mathcal{E}_d=0.1$, $\tilde{\omega}_0=10$, and $\alpha=0.04$ (see Table~1 in the main text). Panels in (a) and (b) show results for $\textbf{B}_0\parallel [001]$, $\textbf{B}_0\parallel [110]$, and $\textbf{B}_0\parallel [111]$. Black arrows represent the in-plane components, while the out-of-plane component is color-coded.
     }
    \label{fig: skx_polarization}
\end{figure}

Figure~\ref{fig: skx_polarization}(b) shows the spatial profile of the local electric polarization induced by a skyrmion for the three applied magnetic field directions. The local electric polarization exhibits $C_{2z}$ and $C_{3z}$ symmetry for $\textbf{B}_0\parallel [001]$ and $\textbf{B}_0\parallel [111]$, respectively. In contrast, it has no rotational symmetry for $\textbf{B}_0\parallel [110]$, which is responsible for the unidirectional motion of skyrmions under laser irradiation, as discussed in the main text.

\section{Floquet theory for classical spin dynamics}
\label{method: Floquet_theory}

The time-dependent magnetization $\Bm_{\Br}(t)$ can be partitioned into a fast part $\Bn_{\Br}(t+T_0)=\Bn_{\Br}(t)$ and a slow part $\Bm_{\Br,s}(t)$, where $T_0=2\pi/\omega_0$ denotes the period of laser electromagnetic fields.
Given that $\hbar\omega_0$ is much greater than the typical energy of static spin configurations governed by $H_0$, we employ the high frequency expansion of the LLG equation as discussed in Ref.~\cite{higashikawaFloquetEngineeringClassical2018}.

The long-time dynamics $\Bm_{\Br,s}(t)$ is described by an effective LLG equation where the time-dependent effective field $\BH_{\Br}^\textrm{eff}(t)$ in Eq.~\eqref{eq: general_LLG} is replaced by the effective Floquet field $\BH_{\Br,F}$, which up to first order in $1/\omega_0$ is given by
\begin{align}
\BH_{\Br,F}&=\BH_{\Br,F}^{(0)}+\BH_{\Br,F}^{(1)}\nonumber\\
&=\BH_{\Br,0}^\textrm{eff}+\sum_{m\ne 0}\frac{i\gamma[\BH^\textrm{eff}_{\Br,-m},\BH^\textrm{eff}_{\Br,m}]_\textrm{mag}}{2m\omega_0(1+\alpha^2)},
\label{eq: floquet_effective_magnetic}
\end{align}
with 
$\BH^{\textrm{eff}}_{\Br,m}=\frac{1}{T_0}\int^{T_0}_0 dt\, \BH^{\textrm{eff}}_{\Br}(t)e^{i m\omega_0 t}$. The commutator $[\,,\,]_\textrm{mag}$ is defined as
\begin{eqnarray}
&&[\BH^{\textrm{eff}}_{\Br,m},\BH^{\textrm{eff}}_{\Br,n}]_\textrm{mag}\nonumber\\
&\equiv&\BH^{\textrm{eff}}_{\Br,m}\times \BH^{\textrm{eff}}_{\Br,n}+\alpha \boldsymbol{m}_{\Br}\times(\BH^{\textrm{eff}}_{\Br,m}\times \BH^{\textrm{eff}}_{\Br,n}) \nonumber\\
&+&\sum_{\Br'}\Big[(\BH^{\textrm{eff}}_{\Br',m}\cdot \boldsymbol{L}_{\Br'})\BH^{\textrm{eff}}_{\Br,n}-(\BH^{\textrm{eff}}_{\Br',n}\cdot \boldsymbol{L}_{\Br'})\BH^{\textrm{eff}}_{\Br,m}\Big]\nonumber\\
&+&\sum_{\Br'} \alpha\Big[(\boldsymbol{m}_{\Br'}\cdot\BH^{\textrm{eff}}_{\Br',m}\times \boldsymbol{L}_{\Br'} )\BH^{\textrm{eff}}_{\Br,n}\nonumber\\
&-&(\boldsymbol{m}_{\Br'}\cdot\BH^{\textrm{eff}}_{\Br',n}\times \boldsymbol{L}_{\Br'} )\BH^{\textrm{eff}}_{\Br,m}\Big],\quad\quad\quad
\label{eq: magnetic_commutation_def}
\end{eqnarray}
where $\boldsymbol{L}_{\Br}$ is the angular momentum operator whose components are $L_{\Br,a}=-\epsilon_{abc}m_{\Br,b}(\partial/\partial m_{\Br,c})$
%
%
with $\epsilon_{abc}$ denoting the totally antisymmetric tensor of rank 3 and repeated indices are summed over.

The only nonzero Fourier components of the laser electromagnetic fields are written as
\begin{align}
\textbf{E}_{\pm 1}&= \frac{E_d}{2}(\pm i\cos\delta+\sin\delta,1,0),\\
\textbf{B}_{\pm1}&=\frac{B_d}{2}(1,\mp i\cos\delta-\sin\delta,0),
\end{align}
where $\textbf{E}_{m}=\frac{1}{T_0}\int^{T_0}_0dt \,\textbf{E}(t)e^{im\omega_0 t}$ and $\textbf{B}_{m}=\frac{1}{T_0}\int^{T_0}_0dt \,\textbf{B}(t)e^{im\omega_0 t}$.
Using Eq.~\eqref{eq: polarization [110]}, valid for $\textbf{B}_0\parallel [110]$, the corresponding Fourier components of the effective field are given by
\begin{eqnarray}
\BH^\textrm{eff}_{\Br,\pm 1}&=&\textbf{B}_{\pm 1}+\frac{\lambda}{g\mu_B}(-E_{x,\pm 1}m_{\Br,y}-E_{y,\pm 1}m_{\Br,x},\nonumber\\
&-&E_{x,\pm1}m_{\Br,x},E_{y,\pm 1}m_{\Br,z}).
\end{eqnarray}
Therefore, the effective Floquet field up to first order in $1/\omega_0$ is obtained as
\begin{align}
\BH_{\Br,F}^{(1)}&=\frac{i\gamma[\BH^\textrm{eff}_{\Br,-1},\BH^\textrm{eff}_{\Br,+1}]_{\textrm{mag}}}{(1+\alpha^2)\omega_0}\nonumber\\
&=B_F\Big\{ -\mathcal{E}_d m_{\Br,z}(1+4\mathcal{E}_dm_{\Br,x}),2\mathcal{E}_d^2 m_{\Br,y}m_{\Br,z},\nonumber\\
&1-\mathcal{E}_d m_{\Br,x}+\mathcal{E}_d^2(-2m_{\Br,x}^2+m_{\Br,y}^2) \Big\},
\label{eq: effective_field_[110]}
\end{align}
with $B_F=\frac{\gamma B_d^2\cos\delta}{2(1+\alpha^2)\omega_0}$ and $\mathcal{E}_d= \lambda c / g\mu_{B}$.
Since $\lambda$ is generally frequency dependent~\cite{takahashiMagnetoelectricResonanceElectromagnons2012,huvonenMagneticExcitationsOptical2009}, we introduce the dimensionless parameter $\mathcal{E}_d$ for the strength of magnetoelectric coupling in the THz regime.
Noting the reference value of $\mathcal{E}_d=0.092$ for $\lambda=5.64\times 10^{-27}\mu$Cm~\cite{sekiMagnetoelectricNatureSkyrmions2012,mochizukiMagnetoelectricResonancesPredicted2013}, we use $\mathcal{E}_d=0.1$ in the main text and vary it up to $\CalE_d=0.3$ in the Supplementary Material. 
It should also be noted that small contributions proportional to the damping coefficient $\alpha$ are neglected in Eq.~\eqref{eq: effective_field_[110]}, which do not qualitatively affect the result as shown in an excellent agreement between the LLG simulation and effective LLG equation (see Fig~\ref{fig: LLG_vs_Floq}).

The effective Floquet Hamiltonian in Eq.~(4) of the main text is constructed by inverting the relation $\BH_{\Br,F}=-[1/(\hbar \gamma S)]\partial H_F/\partial\Bm_{\Br}$. In Section~\ref{section: effective_H}, we derive the effective Floquet Hamiltonian for $\textbf{B}_0\parallel [001]$ and $\textbf{B}_0\parallel [111]$.


According to~\cite{higashikawaFloquetEngineeringClassical2018}, the time evolution of the non-equilibrium steady state of the magnetization between $t$ and $t + \delta t$ is given by
\begin{equation}
\Bm_{\Br}(t + \delta t) - \Bm_{\Br}(t) = \frac{\boldsymbol{f}_{\Br,\textrm{mic}}^{(1)}(t+\delta t)}{\omega_0} - \frac{\boldsymbol{f}_{\Br,\textrm{mic}}^{(1)}(t)}{\omega_0} + \mathcal{O}(\delta t/\omega_0),  
\end{equation}
where we assumed $\omega_0\gg 1$. In the above expression, the short-time dynamics is included in 
\begin{equation}
    \boldsymbol{f}_{\Br,\textrm{mic}}^{(1)}(t)=- \frac{\gamma \Bm_{\Br}(t)}{(1+\alpha^2)}
  \times\Big[ \BH^{(1)}_{\Br,\mathrm{mic}}(t)+\alpha\Bm_{\Br}(t)\times\BH^{(1)}_{\Br,\mathrm{mic}}(t) \Big],
  \label{eq: SM_microscopic_force}
\end{equation}
where
\begin{equation}
\BH^{(1)}_{\Br,\mathrm{mic}}(t)=-i\sum_{m\ne 0}\frac{\BH_{\Br,-m}^\textrm{eff}e^{im\omega_0 t}}{m}\,,
\label{eq: kick_field}
\end{equation}
which for $\textbf{B}_0\parallel [110]$ takes the form
\begin{align}
\BH^{(1)}_{\Br,\mathrm{mic}}(t)&=-i(\BH^\textrm{eff}_{\Br,-1}e^{i\omega_0 t}-\BH^\textrm{eff}_{\Br,+1}e^{-i\omega_0 t})\nonumber\\
&=B_d\Big\{ (1-\mathcal{E}_d m_{\Br,x})\sin \,(\omega_0 t)\nonumber\\
&+\mathcal{E}_d m_{\Br,y}\cos\,(\omega_0 t+\delta),\nonumber\\
&(1+\mathcal{E}_d m_{\Br,x})\cos\,(\omega_0 t+\delta), \mathcal{E}_d m_{\Br,z}\sin\,(\omega_0 t)\Big\}.
\label{eq: kicked_[110]}
\end{align}
When the system is in the steady state, we can write $\Bm_{\Br}(t) = \Bm_{\Br,s} +\Bn_{\Br}(t)$, hence
\begin{equation}
\Bn_{\Br}(t+\delta t) - \Bn_{\Br}(t) = \frac{\boldsymbol{f}_{\Br,\textrm{mic}}^{(1)}(t+\delta t)}{\omega_0}-\frac{\boldsymbol{f}_{\Br,\textrm{mic}}^{(1)}(t)}{\omega_0} + \mathcal{O}(\delta t/\omega_0)\,.
\end{equation}
Taking the limit $\delta t\rightarrow 0$ we obtain
\begin{equation}
\frac{d\Bn_{\Br}(t)}{dt}=\frac{1}{\omega_0}\frac{d\boldsymbol{f}_{\Br,\textrm{mic}}^{(1)}(t)}{dt}.
\end{equation}
Since any constant value of the magnetization is part of $ \Bm_{\Br,s}$, upon integrating, we get to leading order in $1/\omega_0$ 
\begin{eqnarray}
    \Bn_{\Br}(t)&=&\frac{\boldsymbol{f}_{\Br,\mathrm{mic}}^{(1)}(t)}{\omega_0}.
    \label{eq:oscillating}
\end{eqnarray}

\section{The effective Floquet Hamiltonian for $\textbf{B}_0\parallel [001]$ and $\textbf{B}_0\parallel[111]$ }
\label{section: effective_H}
As discussed in Section~\ref{section: multiferroics}, the local electric polarization of Cu$_2$OSeO$_3$ takes different expressions depending on the direction of the applied magnetic field $ \textbf{B}_0$.
In the following, we provide expressions for the effective Floquet field $\BH_{\Br,F}^{(1)}$, the effective Floquet Hamiltonian $H_F$, and the drift field $\BH^{(1)}_{\Br,\mathrm{mic}}$ for $\textbf{B}_0\parallel [001]$ and $\textbf{B}_0\parallel[111]$, which can be derived analogously to those corresponding to $\textbf{B}_0\parallel [110]$.

The effective Floquet fields are obtained as
\begin{align}
\BH_{\Br,F}^{(1)}&=B_F\big\{- 2\mathcal{E}_dm_{\Br,y}+2\mathcal{E}_d^2 m_{\Br,x}m_{\Br,z},\nonumber\\
&-2\mathcal{E}_dm_{\Br,x}+2\mathcal{E}_d^2 m_{\Br,y}m_{\Br,z},\nonumber\\
&\quad 1+\mathcal{E}_d^2(m_{\Br,x}^2+m_{\Br,y}^2-3m_{\Br,z}^2) \big\}\textrm{ for }\textbf{B}_0\parallel [001],
\end{align}
and
\begin{align}
\BH_{\Br,F}^{(1)}&=B_F\Big\{2\mathcal{E}_d^2 m_{\Br,x}(\sqrt{2}m_{\Br,y}-m_{\Br,z}),\nonumber\\
&\sqrt{2} \mathcal{E}_d^2m_{\Br,x}^2-\mathcal{E}_d^2m_{\Br,y}(\sqrt{2}m_{\Br,y}+2m_{\Br,z}),\nonumber\\
&\quad 1-\mathcal{E}_d^2(m_{\Br,x}^2+m_{\Br,y}^2-m_{\Br,z}^2) \Big\} 
\textrm{ for }\textbf{B}_0\parallel [111],
\end{align}
with $B_{F}=\frac{\gamma B_d^2\cos\delta }{2(1+\alpha^2)\omega_0}$.
The corresponding effective Floquet Hamiltonians are obtained as
\begin{align}
H_F&=H_0+\sum_{\Br}g\mu_BB_{F}\Big[-\hat{z}\cdot \Bm_{\Br}+ 2\mathcal{E}_dm_{\Br,x}m_{\Br,y}\nonumber\\
&+\mathcal{E}_d^2m_{\Br,z}(-m_{\Br,x}^2-m_{\Br,y}^2+m_{\Br,z}^2)\Big]\textrm{ for }\textbf{B}_0\parallel [001] ,
\label{eq: effective_H_[001]}
\end{align}
and 
\begin{align}
H_F&=H_0+\sum_{\Br}g\mu_B B_{F}\Big[\sqrt{2}\mathcal{E}_d^2m_{\Br,y}(-m_{\Br,x}^2+\frac{m_{\Br,y}^2}{3})\nonumber\\
&-\hat{z}\cdot \Bm_{\Br}+\mathcal{E}_d^2m_{\Br,z}\Big(m_{\Br,x}^2+m_{\Br,y}^2-\frac{m_{\Br,z}^2}{3}\Big)\Big],\nonumber\\
&\textrm{ for }\textbf{B}_0\parallel [111] .
\label{eq: effective_H_[111]}
\end{align}
The drift fields $ \BH_{\Br,\textrm{mic}}^{(1)}(t)$ are obtained from Eq.~(16) of the main text as 
\begin{align}
\BH_{\Br\textrm{,mic}}^{(1)}(t)&=B_d\Big\{\sin \,(\omega_0 t)+\mathcal{E}_d m_{\Br,z}\cos\, (\omega_0 t+\delta),\nonumber\\
&\cos \,(\omega_0 t+\delta)+\mathcal{E}_d m_{\Br,z}\sin\, (\omega_0 t), \nonumber\\
&\quad \mathcal{E}_d m_{\Br,x}\cos\,(\omega_0 t+\delta)+\mathcal{E}_dm_{\Br,y}\sin\,(\omega_0 t)\Big\}\nonumber\\
&\quad\textrm{ for } \textbf{B}_0\parallel [001],
\end{align}
and
\begin{align}
\BH_{\Br\textrm{,mic}}^{(1)}(t)&=\frac{B_d}{\sqrt{6}}\Big\{ (\sqrt{6}-2\mathcal{E}_d m_{\Br,x})\sin \,(\omega_0 t)\nonumber\\
&+\mathcal{E}_d  (2m_{\Br,y}+\sqrt{2}m_{\Br,z})\cos\,(\omega_0 t+\delta),\nonumber\\
&\quad(\sqrt{6}+2\mathcal{E}_d m_{\Br,x})\cos\,(\omega_0 t+\delta)\nonumber\\
&+\mathcal{E}_d(2m_{\Br,y}-\sqrt{2}m_{\Br,z})\sin\,(\omega_0 t), \nonumber\\
&\quad \sqrt{2}\mathcal{E}_d  m_{\Br,x}\cos\,(\omega_0 t+\delta)-\sqrt{2}\mathcal{E}_dm_{\Br,y}\sin\,(\omega_0 t)\Big\} \nonumber\\
&\quad\textrm{ for } \textbf{B}_0\parallel [111].
\end{align}

We now comment on the distortion of the spin texture by the effective Floquet Hamiltonian $H_F$.
In Fig~\ref{fig: skx_polarization}(c), we show the distortion $\delta\Bm_{\Br}$ in the spin textures of a single skyrmion under left-circularly polarized laser (LCP) irradiation, defined as 
\begin{equation}
\delta\Bm_{\Br}=\Bm_{\Br}(\CalE_d=0.1)-\Bm_{\Br}(\CalE_d=0),
\end{equation}
where $\Bm_{\Br}(\CalE_d)$ represents the magnetic configuration obtained by solving the effective LLG equation with magnetoelectric coupling strength $\CalE_d$.
We note that the spin texture $\Bm_{\Br}(\CalE_d)$ is equivalent to the time-averaged spin texture under laser irradiation.
As shown in the middle panel of Fig.~\ref{fig: skx_polarization}(c), the dipole-like distortion is observed in $\delta\Bm_{\Br}$ for $\textbf{B}_0\parallel [110]$, exhibiting a similar feature to the local electric polarization.
Equivalent results are obtained for $\textbf{B}_0\parallel [001], [111]$ in Fig.~\ref{fig: skx_polarization}(c), showing a quadrupole-like feature and a ring pattern, respectively.
Therefore, the magnetoelectric coupling results in the imprinting of local electric multipole moments on the time-averaged spin textures of skyrmions under LCP light.  
We should note that the rotational symmetry of $\delta\Bm_{\Br}$ is equivalent to the local electric polarization with $C_{2z}/C_{3z}$ symmetry for $\textbf{B}_0\parallel [001]/[111]$.
For a right-circularly polarized laser (RCP), the induced distortion patterns are reversed due to the change in the sign of $B_F$.


\section{Velocity of laser-driven skyrmion motion}

In order to employ Thiele's approach~\cite{thieleSteadyStateMotionMagnetic1973} to obtain an expression for the drift velocity $\boldsymbol{v}$ of skyrmions, we first take the continuum limit $\Bm_{\Br}(t)\rightarrow \Bm(\Br,t)$. Assuming the non-equilibrium steady state has already been reached, we can write the magnetization as $\Bm(\Br,t) = \Bm_\textrm{ave}(\Br)+\Bn(\Br,t)$, where $\Bm_\textrm{ave}(\Br)=\braket{\Bm(\Br,t)}_{T_0}$ denotes the time-averaged magnetization with $\braket{f}_{T_0}=\frac{1}{T_0}\int^{T_0}_0 dt f(t)$.
Then the net force exerted on a skyrmion is given as~\cite{wangDrivingMagneticSkyrmions2015}    
\begin{align}
\textrm{F}_i&=-\gamma \int \Bm_\textrm{ave}(\Br)\cdot \Big[\partial_i \Bm_\textrm{ave}(\Br)\nonumber\\
&\times \Braket{\Bm(\Br,t)\times \boldsymbol{H}^\textrm{eff}(\Bm(\Br,t),t)}_{T_0}\Big]dx dy,
\label{eq: thiele_netforce}
\end{align}
with $i=x,y$.
We neglect distortions due to the laser fields and assume a rotationally symmetric texture 
\begin{equation}
\Bm_\textrm{ave}(\Br)=\{\cos \phi(\psi)\sin\theta(r),\sin \phi(\psi)\sin\theta(r),\cos\theta(r)\},
\end{equation}
with $\Br=\{r\cos\psi, r\sin\psi \}$. The oscillating term $\Bn(\Br,t)$ is obtained from Eq.~\eqref{eq: SM_microscopic_force} and \eqref{eq:oscillating} by replacing $\Bm_{\Br}(t)$ with $\Bm_\textrm{ave}(\Br)$. 

We consider a general expression for the local electric polarization, quadratic in spin components
\begin{equation}
P_{r,i}=\frac{g\mu_B}{c}\chi^i_{jk}m_{\Br,j}m_{\Br,k},
\end{equation}
with $i,j,k=x,y,z$ and $c$ denotes the speed of light.
Up to first order in the Floquet-Magnus expansion, the oscillatory part $\Bn(\Br,t)$ is obtained as
\begin{align}
    \Bn(\Br,t)&=- \frac{\gamma \Bm_\textrm{ave}(\Br)}{(1+\alpha^2)\omega_0} \times\Big[ \BH^{(1)}_{\mathrm{mic}}(\Br,t)\nonumber\\
    &+\alpha\Bm_\textrm{ave}(\Br)\times\BH^{(1)}_{\mathrm{mic}}(\Br,t) \Big],
    \label{eq: n(t)}
\end{align}
where 
\begin{align}
    [\BH^{(1)}_{\mathrm{mic}}(\Br,t)]_j&=-i(\BH^\textrm{eff}_{-1}(\Br)e^{i\omega_0 t}-\BH^\textrm{eff}_{+1}(\Br)e^{-i\omega_0 t})_j\nonumber\\
    &=-i\Big\{\Big(B_{- 1,j}+\frac{\chi^i_{jk}E_{- 1,i}m_{\textrm{ave},k}(\Br)}{c}\Big)e^{i\omega_0 t}\nonumber\\
    &\quad -\textrm{c.c.}\Big\},
\end{align}
with $\BH^\textrm{eff}_{m}(\Br)=\frac{1}{T_0}\int^{T_0}_0dt \,\BH^\textrm{eff}(\Bm(\Br,t),t)e^{im\omega_0 t}$, $\textbf{E}_{m}=\frac{1}{T_0}\int^{T_0}_0dt \,\textbf{E}(t)e^{im\omega_0 t}$, and $\textbf{B}_{m}=\frac{1}{T_0}\int^{T_0}_0dt \,\textbf{B}(t)e^{im\omega_0 t}$.
It is straightforward to show that the azimuthal integration of $\braket{\Bm_\textrm{ave}(\Br)\times \BH^\textrm{eff}(\Bm_\textrm{ave}(\Br),t)}_{T_0}$ vanishes.
Hence, the lowest order of $\textbf{F}$ in the high frequency expansion is given by collecting linear terms in $\Bn(\Br,t)$, which are obtained as
\begin{align}
&\braket{\{\Bm_\textrm{ave}(\Br)+\Bn(\Br,t)\}\times \BH^\textrm{eff}(\Bm(\Br,t),t)}_{T_0}\nonumber\\
&=\Bm_\textrm{ave}(\Br)\times \BH_{0}^{\textrm{eff}}(\Br)+\Bn_{\pm1}(\Br)\times \BH_{\mp 1}^{\textrm{eff}}(\Br),
\end{align}
where
\begin{align}
[\Bm_\textrm{ave}(\Br)\times\BH_{0}^{\textrm{eff}}(\Br)]_a&=\epsilon_{abc} m_{\textrm{ave},b}(\Br)\frac{E_{\pm1,i}}{c}\chi^i_{ck}n_{\mp 1,k}(\Br),\\
[\Bn_{\pm1}(\Br)\times \BH^\textrm{eff}_{\mp 1}(\Br)]_a&=\epsilon_{abc}n_{\pm1,b}(\Br)(B_{\mp1,c}\nonumber\\
&+\frac{E_{\mp1,i}}{c}\chi^i_{ck}m_{\textrm{ave},k}(\Br)).
\end{align}
with $a=x,y,z$.

Evaluating the azimuthal angular integration with $\phi=\psi-\pi/2$ for Bloch skyrmions, we find that zeroth order terms in $\alpha$ cancel each other out.
Thus the net force $\textbf{F}$ is proportional to $\alpha$ and given by
\begin{align}
\textrm{F}_x&=\frac{\pi\gamma B^2_d \alpha \cos\delta}{12(1+\alpha^2)\omega_0}\int^\infty_0 dr \Big[4(\chi^x_{xy}-2\chi^y_{xx}+2\chi^y_{zz})\sin^3\theta\nonumber\\
&-3(\chi^x_{zx}\chi^y_{xy}-\chi^x_{xy}\chi^y_{zx}+2\chi^x_{yz}\chi^y_{yy}-2\chi^x_{yy}\chi^y_{yz}\nonumber\\
&+2\chi^x_{zz}\chi^y_{yz}-2\chi^x_{yz}\chi^y_{zz})\Big(2\frac{d\theta}{dr} r+\sin 2\theta\Big)\Big],\nonumber\\
\textrm{F}_y&=-\frac{\pi \gamma B_d^2\alpha\cos\delta}{12(1+\alpha^2)\omega_0}\int^\infty_0 dr \Big[4(\chi^y_{xy}-2\chi^x_{yy}+2\chi^x_{zz})\sin^3\theta\nonumber\\
&-3(\chi^x_{yz}\chi^y_{xy}-\chi^x_{xy}\chi^y_{yz}+2\chi^x_{zx}\chi^y_{xx}-2\chi^x_{xx}\chi^y_{zx}\nonumber\\
&+2\chi^x_{zz}\chi^y_{zx}-2\chi^x_{zx}\chi^y_{zz})\Big(2\frac{d\theta}{dr} r+\sin 2\theta\Big)\Big].
\end{align}
For $\textbf{B}_0\parallel [110]$, we substitute $\chi^x_{xy}=-\chi^z_{yz}=-\mathcal{E}_d$ and $\chi^y_{xx}=- \chi^y_{zz}=-\mathcal{E}_d/2$ with $\mathcal{E}_d=c\lambda/g\mu_B $. In this case, the total force is given by
\begin{align}
\textrm{F}_x&=\frac{\pi\gamma B^2_d\mathcal{E}_d\alpha \cos\delta}{3(1+\alpha^2)\omega_0}\int^\infty_0 dr \sin^3\theta, \quad F_y=0.\label{eq: F_110}
\end{align}
Finally, the velocity of laser-driven skyrmions is derived as
\begin{align}
    v_x&= \frac{\alpha\eta \textrm{F}_x}{(4\pi Q)^2+\alpha^2\eta^2}\approx \frac{\alpha\eta \textrm{F}_x}{(4\pi Q)^2}=-\frac{\alpha\eta}{4\pi Q} v_0\overline{B}_F\CalE_d\alpha C,\nonumber\\
v_y&=\frac{-4\pi Q\textrm{F}_x}{(4\pi Q)^2+\alpha^2\eta^2}\approx-\frac{F_x}{4\pi Q}=v_0\overline{B}_F\CalE_d\alpha C,
\label{eq: SM_velocity_formula}
\end{align}
where $\eta=\int |\partial_x \Bm_\textrm{ave}(\Br)|^2 dxdy\approx 4\pi$ depends on the spatial profile of skyrmions, $Q$ is the topological charge, $v_0=a\gamma  D^2/g\mu_BJ\approx 6.16$ m/s is a characteristic velocity, $\overline{B}_F=g\mu_B B_FJ/D^2$, and the constant $C$ is evaluated as $C\approx -2.25/Q$ by extracting $\theta(\overline{r})$ from the numerically obtained skyrmion configuration in Fig.~\ref{fig: skx_polarization}. 
In contrast, substituting $\chi^{x}_{zx}=-\chi^y_{yz}=-\mathcal{E}_d$ and $\chi^{z}_{xx}=-\chi^{z}_{yy}=-\mathcal{E}_d/2$ for $\textbf{B}_0\parallel [001]$, we find $\textbf{F}=0$.
Similarly, it is straightforward to show that $\textbf{F}=0$ for $\textbf{B}_0\parallel [111]$.

There is a strong analogy between laser-driven skyrmions and deterministic ratchets~\cite{mateosChaoticTransportCurrent2000, zapataDeterministicRatchetStationary2009}, both of which require asymmetry and nonlinearity for unidirectional motion.
Since the local electric polarization breaks the rotational symmetry of skyrmions, the magnetoelectric effect introduces asymmetry in the system~\cite{liuSkyrmionDynamicsMultiferroic2013}. 
This is in stark contrast with setups for skyrmion motion driven by microwave fields~\cite{wangDrivingMagneticSkyrmions2015, ikkaResonanceModesMicrowavedriven2018, takeuchiSelectiveActivationIsolated2018, yuanWigglingSkyrmionPropagation2019}, where the rotational symmetry is explicitly broken by applying in-plane magnetic fields. In comparison to previous proposals using microwave fields, another advantage of our work is that there is no need to fine-tune the laser frequency, hence simplifying the experimental setup. We also find that the nonlinear Gilbert damping effect is fundamental to drive skyrmions by THz lasers. While the skyrmion motion is usually suppressed by the effect of damping at resonance~\cite{yuanWigglingSkyrmionPropagation2019}, the velocity of skyrmions driven by THz lasers is enhanced with increasing damping constant as discussed in Section~\ref{section: numerical_velocity}.
Beyond the phenomenological Gilbert damping considered in this work, we should note that the damping kernel derived from intrinsic skyrmion magnon modes was shown to drive skyrmions under oscillating magnetic field gradients~\cite{psaroudakiSkyrmionsDrivenIntrinsic2018}.

\section{Numerical study of laser-driven velocity}
\label{section: numerical_velocity}
\begin{figure}[t]
    \centering
    \includegraphics[width=\columnwidth]{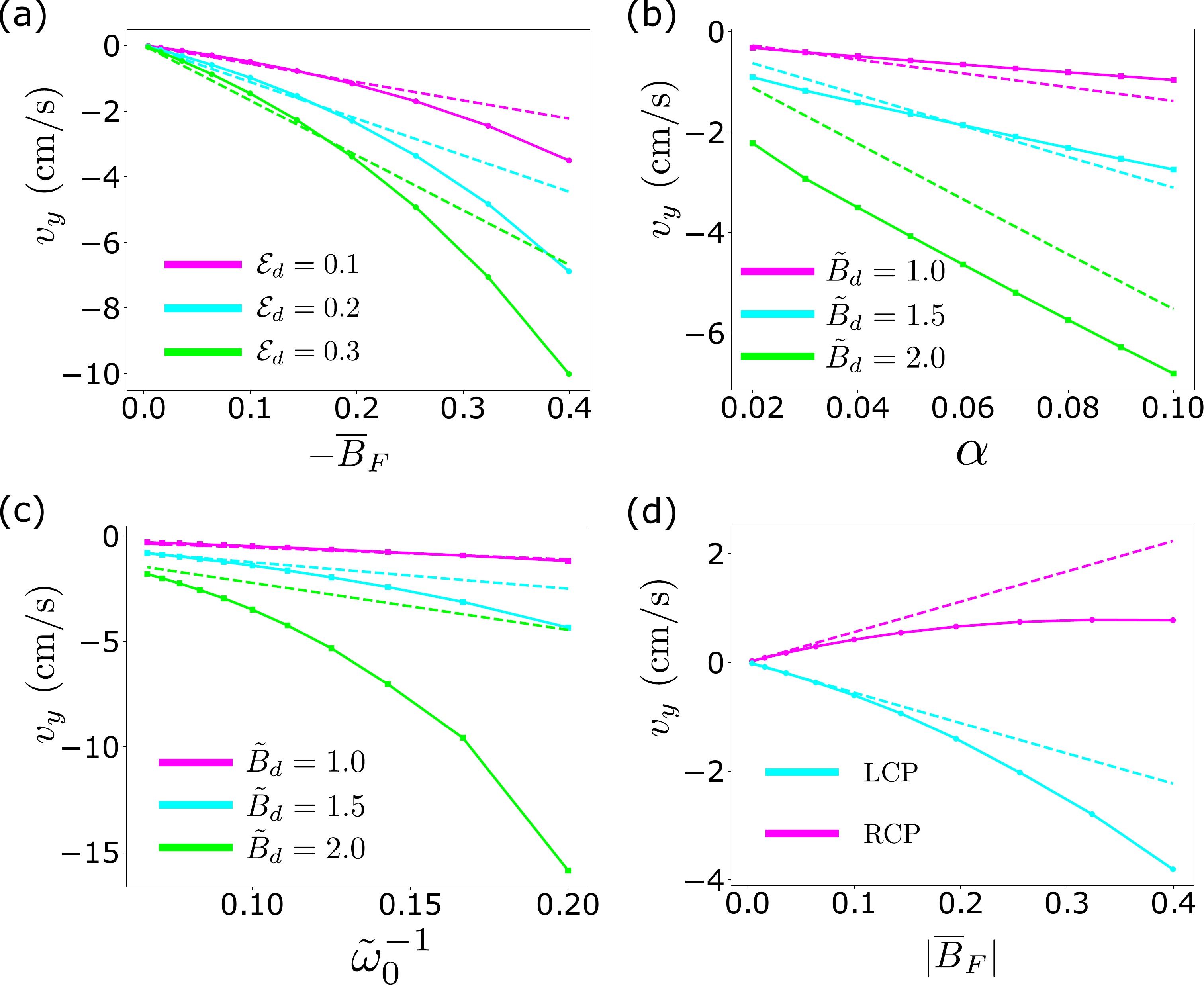}
    \caption{\textbf{Parameter dependence of laser-driven skyrmion velocity for  $\textbf{B}_0\parallel[110]$.} (a)-(c) Skyrmion velocity as a function of (a) the effective magnetic field $\overline{B}_F\propto B_d^2$, (b) the phenomenological damping constant $\alpha$, (c) the frequency of applied laser fields $\tilde{\omega}_0$. The initial configuration is the same as in Fig.~\ref{fig: skx_polarization}(a), obtained at $\overline{B}_0=0.9$.
    (d) Velocity of skyrmion crystals under RCP and LCP light as a function of $|\overline{B}_F|\propto B_d^2$. The spin configuration of skyrmion crystals is prepared at $\overline{B}_0=0.4$. 
    The parameters are fixed as $D/J=0.09$, $\tilde{\omega}_0=10$, $\mathcal{E}_d=0.1$, and $\alpha=0.04$, unless otherwise stated in each panel. Solid and dashed lines indicate the result of LLG simulations and the analytical expression of $v_y$ obtained in the main text, respectively.
    }
    \label{fig: velocity1}
\end{figure}

We investigate the parameter dependence of the laser-driven velocity $v_y$ of both single skyrmions and skyrmion crystals for $\textbf{B}_0\parallel[110]$. 
We solve the LLG equation for a total duration of $\overline{t}=50,000$ (the time step is $\Delta\overline{t}=10^{-2}$). 
To ensure that the system is in a steady state, only the final time interval of duration $\overline{t}=30,000$ is considered for the velocity calculation. 
Firstly, we compute $v_y$ of single skyrmions under LCP light. The initial spin configuration of a single skyrmion is prepared by Monte Carlo simulations with 150$\times$150 spins under periodic boundary conditions at $D/J=0.09$ and $\overline{B}_0=0.9$ in Figs.~\ref{fig: velocity1}(a)-(c).

Figures~\ref{fig: velocity1}(a)-(c) show the dependence of $v_y$ on (a) the effective magnetic field $\overline{B}_F=g\mu_B B_FJ/D^2\propto B_d^2$, (b) the damping constant $\alpha$, and (c) the inverse of the driving frequency $\tilde{\omega}_0^{\,-1}$.
Solid and dashed lines are obtained by LLG simulations and the analytical expression of Eq.~(5) in the main text, respectively.
For small $\overline{B}_F$, there is a remarkable agreement between the numerical and analytical results in Fig.~\ref{fig: velocity1}(a).
As we increase the effective field $\overline{B}_F$, their difference becomes larger due to higher order terms in the Floquet-Magnus expansion.
We also find that $v_y$ is proportional to the magnetoelectric coupling strength $\CalE_d$, as predicted in Eq.~(5) of the main text.
Similarly, we find a linear dependence in $\alpha$ and $\tilde{\omega}_0^{\,-1}$ in Figs.~\ref{fig: velocity1}(b)-(c), although higher order contributions result in discrepancies for larger values of $\tilde{B}_d$.
In Fig.~\ref{fig: velocity1}(b), there is a constant shift of $v_y$ in the numerical result compared to the analytical result for $\tilde{B}_d=2.0$, implying that $\alpha$-independent terms in $v_y$ might arise from higher order terms in the Floquet-Magnus expansion.
In our simulation, the maximum velocity was obtained as $v_y=-15.9$ cm/s at $\alpha=0.04$, $\tilde{\omega}_0=5$, and $\tilde{B}_d=2.0$, but an even larger velocity is possible by tuning $\tilde{\omega}_0$ and $\tilde{B}_d$.

Secondly, we consider the laser-driven motion of skyrmion crystals. The spin configuration of skyrmion crystals is prepared at $D/J=0.09$ and $\overline{B}_0=0.4$ using Monte Carlo annealing with 144$\times$96 spins under periodic boundary conditions.
Following the same procedure as above, we study the velocity of skyrmion crystals under right-circularly polarized (RCP) as well as LCP light as a function of $\overline{B}_F$.
The other parameters are fixed as $\CalE_d=0.1$, $\alpha=0.04$, and $\tilde{\omega}_0=10$.
Figure~\ref{fig: velocity1}(d) shows that the sign of $v_y$ is inverted between RCP and LCP light, as the sign of the effective field $\overline{B}_F$ changes with the chirality of the laser.
Although Eq.~(5) in the main text was derived for a single skyrmion, it shows an excellent agreement with the numerical results for skyrmion crystals at small $|\overline{B}_F|$. 
Interestingly, $v_y$ for RCP light becomes smaller than the analytic result for larger $|\overline{B}_F|$, while it is more enhanced for LCP light. This implies the existence of a negative quadratic contribution in $\overline{B}_F$ regardless of the laser chirality.

\begin{figure}[t]
    \centering
    \includegraphics[width=\columnwidth]{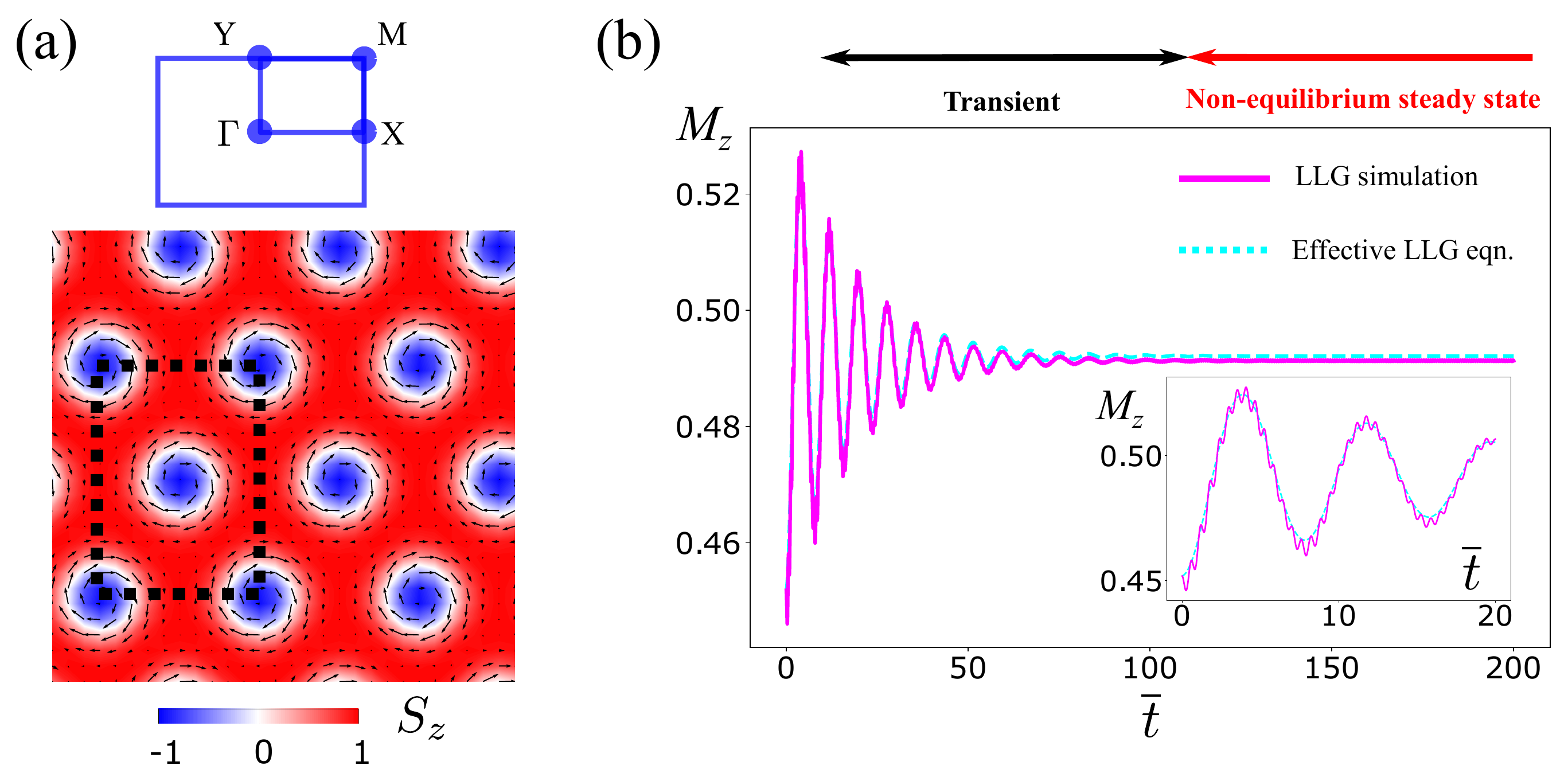}
    \caption{ \textbf{Comparison between LLG and Floquet theory.} (a) Classical ground-state spin texture of a skyrmion crystal at $D/J=1.0$ and $\overline{B}_0=0.4$. The magnetic unit cell is indicated by black dashed lines with the corresponding first Brillouin zone depicted in blue (inset). (b) Time evolution of the out-of-plane component of the total magnetization per site under RCP light obtained by LLG simulations (solid magenta) and the effective LLG equation (dashed cyan) for $\textbf{B}_0\parallel[110]$.
    Parameters are taken as $D/J=1.0$, $\overline{B}_0=0.4$, $\tilde{B}_d=1.0$, $\mathcal{E}_d=0.1$, $\tilde{\omega}_0=10$, and $\alpha=0.04$. The inset shows the result for a shorter time scale, illustrating oscillations at the driving frequency $\omega_0$.
    }
    \label{fig: LLG_vs_Floq}
\end{figure}

\section{Spin wave excitations from the non-equilibrium steady state}
In this section, we discuss low energy spin wave excitations from the non-equilibrium steady state (NESS) of skyrmion crystals under circularly polarized laser irradiation.
Although we consider only $\textbf{B}_0\parallel [110]$ in the following, the resonance frequencies of spin wave modes are dominantly affected by the laser-induced effective magnetic field.
Hence, we expect that our result also applies to $\textbf{B}_0\parallel [001]$ and $\textbf{B}_0\parallel [111]$.
Nonreciprocal magnon band structures induced by the magnetoelectric coupling are discussed in Section \ref{section: edge states}.
To decrease the computational time here and in the subsequent sections, we reduce the size of the magnetic unit cell by choosing $D/J=1.0$.
The classical ground-state spin textures of skyrmion crystals are obtained by Monte Carlo simulated annealing at $\overline{B}_0=0.4$ for 30$\times $30 spin systems as shown in Fig.~\ref{fig: LLG_vs_Floq}(a).
The LLG equation is then solved for the skyrmion crystal under RCP light.

Figure~\ref{fig: LLG_vs_Floq}(b) shows the time evolution of the out-of-plane component of the total magnetization per site $M_z(t)=\sum_{\Br}m_{\Br,z}(t)/N_s$ with $N_s$ denoting the total number of spins (solid magenta).
We find large oscillations in $M_z$ corresponding to an initial transient regime. 
The frequency of oscillation in this regime is much smaller than the laser frequency, as indicated by the inset of Fig.~\ref{fig: LLG_vs_Floq}(b).
Thus, the laser frequency is decoupled from resonance frequencies of the magnetic system.
At $\overline{t}\sim100$ the system reaches a NESS, whose time-averaged spin texture corresponds to the static solution of the effective LLG equation. 
In Figure~\ref{fig: LLG_vs_Floq}(b), we also plot the time evolution of $M_z$ obtained by solving the effective LLG equation (dashed cyan).
The result from the effective LLG equation shows an excellent agreement with the LLG simulation.
The difference is that the result from the LLG simulation shows oscillations at the laser frequency $\omega_0$ about the time-averaged magnetization obtained by the effective LLG equation, which are due to the effect of the fast oscillatory part $\Bn(\Br,t)$.

\begin{figure}[t]
    \centering
    \includegraphics[width=\columnwidth]{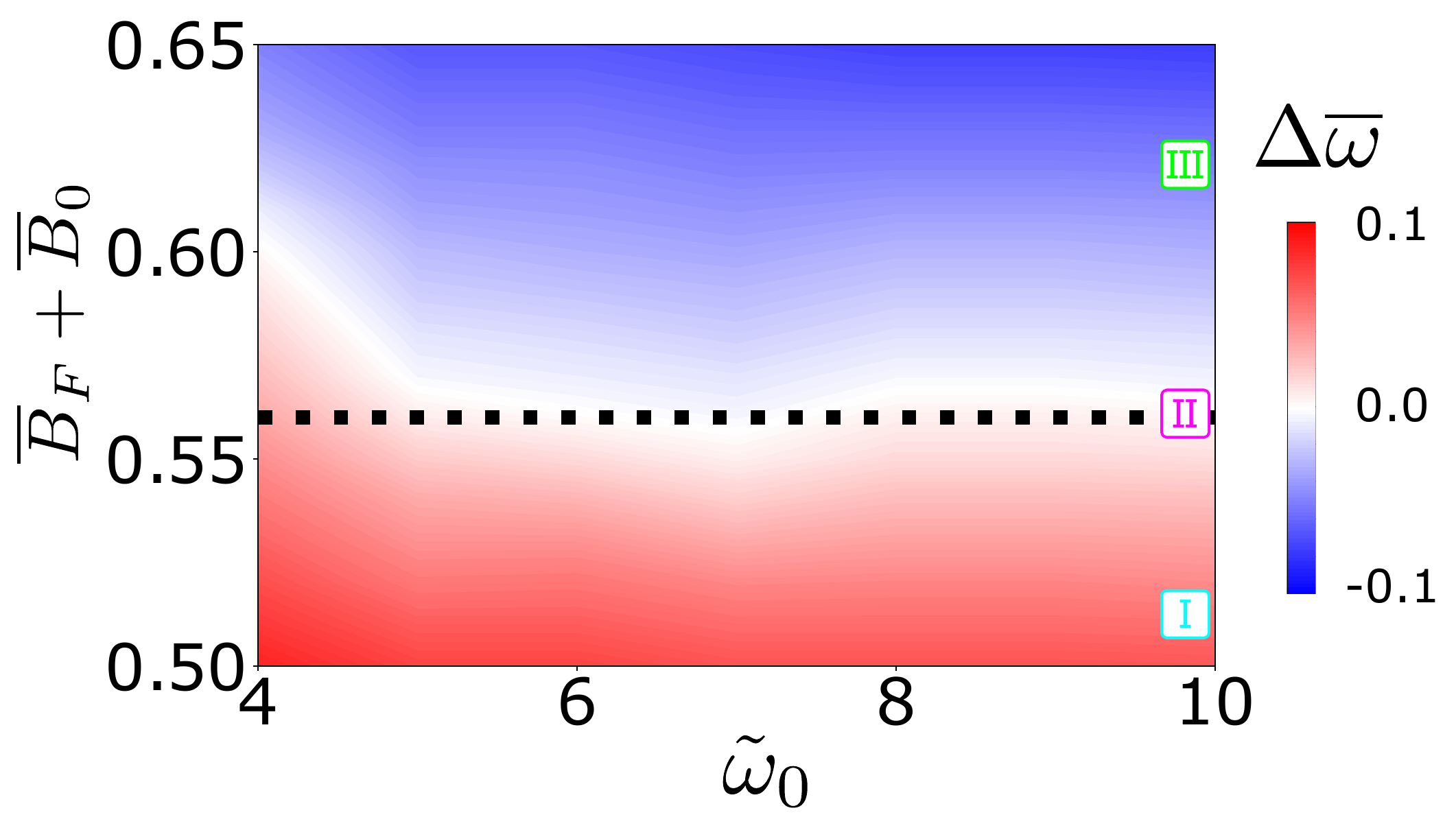}
    \caption{ \textbf{Breakdown of the Floquet-Magnus expansion at low frequencies.} The band gap between the counterclockwise and breathing modes $\Delta\overline{\omega}$ is plotted as a function of the frequency of the applied field $\tilde{\omega}_0$ and the total effective magnetic field $\overline{B}_F+\overline{B}_0$. The dashed line indicates the critical magnetic field $\overline{B}_F+\overline{B}_0=0.56$ for $\tilde{\omega}_0=10$. The parameters are $D/J = 1.0$, $\overline{B}_0 = 0.4$, $\mathcal{E}_d = 0.1$, and $\alpha = 0.04$. Roman numerals indicate the parameters chosen for Fig.~3 in the main text.
    }
    \label{fig: delta_vs_omega}
\end{figure}

Following Ref.~\cite{mochizukiSpinWaveModesTheir2012}, we introduce a pulse field $\textbf{B}_\textrm{pulse}=3B_d(\hat{x}+\hat{z})\delta(t-t_0)$.
Here, the relaxation time $\overline{t}_0=\hbar t_0/JS=3000$ is taken sufficiently long to ensure that the system reaches a NESS (see Fig.~\ref{fig: LLG_vs_Floq}).
By applying the pulse field along both in-plane and out-of-plane directions, we excite the magnetically active modes of skyrmions known as counterclockwise (CCW), clockwise (CW), and breathing modes~\cite{mochizukiSpinWaveModesTheir2012}.
Due to a large difference in frequencies, these low energy magnetic excitations can be distinguished from the forced oscillations at $\omega_0$ as shown in the main text. 

In the main text, we show that the band inversion between CCW and breathing modes occurs at the critical effective magnetic field $\overline{B}_F+\overline{B}_0\approx 0.56$ for $\tilde{\omega}_0=10$.
Here, we study the dependence of the critical magnetic field on the laser frequency $\tilde{\omega}_0$.
Since contributions of higher order terms in the Floquet-Magnus expansion become larger as $\tilde{\omega}_0\rightarrow 1$, the effective magnetic field $\overline{B}_F+\overline{B}_0$ at the band inversion should deviate from the critical value at $\tilde{\omega}_0=10$ for small $\tilde{\omega}_0$.
Figure~\ref{fig: delta_vs_omega} shows the band gap $\Delta\overline{\omega}$ between CCW and breathing mode as a function of $\tilde{\omega}_0$ and $\overline{B}_F+\overline{B}_0$.
Intriguingly, the critical value of the effective magnetic field, where $\Delta\overline{\omega}=0$, is roughly constant up to $\tilde{\omega}_0=5$.
This implies that the first order Floquet-Magnus expansion is a good approximation for $\tilde{\omega}_0\ge 5$.
For $\tilde{\omega}_0\le 3$, we find that the system becomes unstable without reaching a NESS.

\section{Floquet magnon Hamiltonian}
Previously, Floquet topological magnons were discussed in collinear ferromagnets and antiferromagnets driven by a laser field which couples to the magnetic order via the Aharonov-Casher effect~\cite{aharonovTopologicalQuantumEffects1984a, owerreFloquetTopologicalMagnons2017, owerreMagnonicFloquetHofstadter2018, nakataLaserControlMagnonic2019, elyasiTopologicallyNontrivialMagnonic2019}. However, this approach is unsuitable for noncollinear magnetic textures whose magnetic unit cell, of paramount importance for the proper description of magnons, may itself become time-dependent~\cite{delserArchimedeanScrewDriven2020}. Furthermore, the energy injection from a periodic drive inevitably leads to a thermal state at long times, restricting the range of validity of the standard Floquet theory of quantum mechanical systems, which does not account for dissipative effects, to a finite, prethermal time interval~\cite{okaFloquetEngineeringQuantum2019}.

In contrast, we treat the time-averaged magnetic configuration of the NESS as the classical spin ground-state, which can be obtained by solving the effective LLG equation. The Floquet magnon Hamiltonian is then obtained by the Holstein-Primakoff (HP) transformation in the same way as in static systems~\cite{diazTopologicalMagnonsEdge2019,diazChiralMagnonicEdge2020}. This approach naturally includes the damping effect that stabilizes a non-equilibrium steady state, whose magnetic excitations support Floquet magnons.


We perform the HP transformation on the effective Floquet Hamiltonian in Eq.~(4) of the main text for $\textbf{B}_0\parallel[110]$.
Firstly, we define spin operators with respect to the local orthonormal basis $(\Be_{\Br}^{1},\Be_{\Br}^{2},\Bm_{\Br})$, where $\Bm_{\Br}$ is a unit vector parallel to $\BS_{\Br}$ and $\Be_{\Br}^{1}\times\Be_{\Br}^{2} = \Bm_{\Br}$.
In this local basis, the spin operators are written as $\bs{S}_{\Br} = \SS_{\Br}^{1}\Be_{\Br}^{1} + \SS_{\Br}^{2}\Be_{\Br}^{2} + \SS_{\Br}^{3}\Bm_{\Br}$, which are transformed as
\begin{align}
\SS_{\Br}^{+} &= (2S - a^\dagger_{\Br} a_{\Br})^{\frac{1}{2}} a_{\Br},\nonumber\\
\SS_{\Br}^{-} &= a_{\Br}^\dagger(2S - a^\dagger_{\Br} a_{\Br})^{\frac{1}{2}}, \label{eq: HP_transformation} \\
\SS_{\Br}^{3} &= S - a_{\Br}^\dagger a_{\Br}, \nonumber
\end{align}
where $\SS_{\Br}^{\pm} = \SS_{\Br}^{1} \pm i\SS_{\Br}^{2}$, and $a_{\Br}$, $a_{\Br}^\dag$ are the HP bosonic operators.
Assuming $S\gg 1$, the Hamiltonian is expanded as a series in $1/S$ and the Floquet magnon Hamiltonian is constructed from the terms quadratic in bosonic operators. 

The effective Floquet Hamiltonian $H_F$ derived in the main text can be rewritten as
\begin{eqnarray}
H_F&=&H_0+\sum_{\Br}\Big[-g\mu_B B_F\BS_{\Br}\cdot\hat{z}+K_{zx}S_{\Br,z}S_{\Br,x}\nonumber\\
&+&N_{xz}S_{\Br,x}^2S_{\Br,z} +N_{yz}S_{\Br,y}^2S_{\Br,z}\Big],
\end{eqnarray}
with $K_{zx}=g\mu_BB_F\mathcal{E}_d$, $N_{xz}=2g\mu_B B_F\CalE_d^2$, and $N_{yz}=-g\mu_B B_F\CalE_d^2$. Since the spin wave Hamiltonian for the static part $H_0$ was derived previously~\cite{diazTopologicalMagnonsEdge2019,diazChiralMagnonicEdge2020}, we only consider additional terms arising from the high frequency expansion.
It is convenient to perform the HP transformation on the following general expressions.
The term proportional to $\CalE_d$ is generally given as 
\begin{eqnarray}
K_{ab}S_{\Br,a}S_{\Br,b}&=&K_{ab}\Big\{\sum_\alpha(\SS_{\Br}^\alpha\Be_{\Br}^\alpha)\cdot \hat{a}\Big\}\Big\{\sum_\beta(\SS_{\Br}^\beta\Be_{\Br}^\beta)\cdot \hat{b}\Big\}\nonumber\\
&=&K_{ab}\sum_{\alpha,\beta}\SS_{\Br}^\alpha \SS_{\Br}^\beta L_{\Br}^{\alpha a}L_{\Br}^{\beta b},
\end{eqnarray}
with $a,b=x,y,z$ denoting the Cartesian coordinates and $\alpha,\beta=1,2,3$ denoting the local orthogonal basis. Here, we have introduced the tensor $L_{\Br}^{\alpha a}=\Be_{\Br}^\alpha\cdot\hat{a}$. 
Only keeping terms quadratic in bosonic operators, we obtain
\begin{eqnarray}
&&K_{ab}\sum_{\alpha,\beta}\SS_{\Br}^\alpha \SS_{\Br}^\beta L_{\Br}^{\alpha a}L_{\Br}^{\beta b}\nonumber\\
&=&\frac{K_{ab}S}{2}\Big[\Big(L_{\Br}^{1a}L_{\Br}^{1b}+L_{\Br}^{2a}L_{\Br}^{2b}-2L_{\Br}^{3a}L_{\Br}^{3b} \Big)(a_{\Br}^\dagger a_{\Br}+a_{\Br} a_{\Br}^\dagger)\nonumber\\
&&+\Big\{L_{\Br}^{1a}L_{\Br}^{1b}+i(L_{\Br}^{1a}L_{\Br}^{2b}+L_{\Br}^{2a}L_{\Br}^{1b})-L_{\Br}^{2a}L_{\Br}^{2b}\Big\}a_{\Br}^\dagger a_{\Br}^\dagger+\textrm{h.c.}\Big].\nonumber\\
\label{eq: floquet_magnon_identity1}
\end{eqnarray}
Similarly, terms proportional to $\CalE_d^2$ take the following form:
\begin{eqnarray}
N_{ab}S_{\Br,a}^2S_{\Br,b} =N_{ab}\sum_{\alpha,\beta,\gamma}\SS_{\Br}^\alpha \SS_{\Br}^\beta \SS_{\Br}^\gamma L_{\Br}^{\alpha a}L_{\Br}^{\beta a}L_{\Br}^{\gamma b}.
\end{eqnarray}
The terms quadratic in bosonic operators are obtained as
\begin{eqnarray}
&&N_{ab}\sum_{\alpha,\beta,\gamma}\SS_{\Br}^\alpha L_{\Br}^{\alpha a}\SS_{\Br}^\beta L_{\Br}^{\beta a}\SS_{\Br}^\gamma L_{\Br}^{\gamma b}\nonumber\\
&=&\frac{N_{ab}S^2}{2}\Big[(L_{\Br,aab}^{113}+2iL_{\Br,aab}^{123}-L_{\Br,aab}^{223})a_{\Br}^\dagger a_{\Br}^\dagger\nonumber\\
&+&(L_{\Br,aab}^{113}+L_{\Br,aab}^{223}-3L_{\Br,aab}^{333})a_{\Br}^\dagger a_{\Br}+\textrm{h.c.}\Big],
\label{eq: floquet_magnon_identity2}
\end{eqnarray}
where we denote
\begin{align}
L_{\Br,aab}^{113}&=2L_{\Br}^{1a}L_{\Br}^{3a}L_{\Br}^{1b}+(L_{\Br}^{1a})^2L_{\Br}^{3b},\nonumber\\
L_{\Br,aab}^{223}&=2L_{\Br}^{2a}L_{\Br}^{3a}L_{\Br}^{2b}+(L_{\Br}^{2a})^2L_{\Br}^{3b},\nonumber\\
L_{\Br,aab}^{333}&=(L_{\Br}^{3a})^2L_{\Br}^{3b},\nonumber\\
L_{\Br,aab}^{123}&=L_{\Br}^{1a}L_{\Br}^{2a}L_{\Br}^{3b}+L_{\Br}^{3a}L_{\Br}^{1a}L_{\Br}^{2b}+L_{\Br}^{2a}L_{\Br}^{3a}L_{\Br}^{1b}.
\end{align}

Since the magnetic unit cell comprises many spins, we partition the position vector into a Bravais lattice vector $\BR$ and a basis vector $\Br_i$ as $\Br=\BR+\Br_i$.
Then we Fourier transform the bosonic operators according to $a_{\Bk i}=1/\sqrt{N}\sum_{\BR}e^{-\Bk\cdot(\BR+\Br_i)}a_{\Br}$, where $N$ is the number of magnetic unit cells.
Combining the contribution from the static part~\cite{diazChiralMagnonicEdge2020,diazTopologicalMagnonsEdge2019}, we obtain 
\begin{align}
\Hsw = \frac{S}{2}\sum_{\Bk}\psi_{\Bk i}^\dagger H_{ij}(\Bk) \psi_{\Bk j}+\mathcal{E}_0 \,,
\end{align}
with
\begin{align}
 H_{ij}(\Bk) =
\begin{pmatrix}
\Omega_{ij}(\Bk) & \Delta_{ij}(\Bk)\\
\Delta_{ij}^*(-\Bk) &\Omega_{ij}^* (-\Bk)
\end{pmatrix}\,,
\label{eq: Floquet_magnon_hamiltonian}
\end{align}
where $\psi_{\Bk i}=(a_{\Bk i}, a^\dagger_{-\Bk i})^T$ and $\CalE_0=-\frac{1}{2}NS^2\sum_{i} \Lambda_{i}$ for $i=1,\ldots, p$ with $p$ denoting the total number of spins in a magnetic unit cell. 
Each expression is given by 
\begin{align}
\Omega_{ij}(\Bk) &= \delta_{ij}\Lambda_i + \frac{1}{2}[ -J_{ij}(\Bk)\Be_{i}^{+}\cdot\Be_{j}^{-} + \BD_{ij}(\Bk)\cdot\Be_{i}^{+}\times\Be_{j}^{-} ],\nonumber\\
\Delta_{ij}(\Bk) &= \delta_{ij}\Lambda'_i +\frac{1}{2}[ -J_{ij}(\Bk)\Be_{i}^{+}\cdot\Be_{j}^{+} + \BD_{ij}(\Bk)\cdot\Be_{i}^{+}\times\Be_{j}^{+} ],\nonumber\\
\Lambda_{i} (\Bk)&= \sum_{j}[ J_{ij}(\Bk=0)\Bm_{i}\cdot\Bm_{j} - \BD_{ij}(\Bk=0)\cdot\Bm_{i}\times\Bm_{j} ]\nonumber\\
&+ \frac{g\mu_B(B_0+B_F)}{S}\zhat\cdot\Bm_{i}\nonumber\\
&+\sum_{a=x,y}SN_{az}\Big(L_{i,aaz}^{113}+L_{i,aaz}^{223}-3L_{i,aaz}^{333}\Big)\nonumber\\
&+K_{zx}\Big(L_{i}^{1z}L_{i}^{1x}+L_{i}^{2z}L_{i}^{2x}-2L_{i}^{3z}L_{i}^{3x}\Big) ,\\
\Lambda'_i(\Bk)&=\sum_{a=x,y}SN_{az}\Big(L_{i,aaz}^{113}+2iL_{i,aaz}^{123}-L_{i,aaz}^{223}\Big)\nonumber\\
&+K_{zx}\Big\{L_{i}^{1z}L_{i}^{1x}+i(L_{i}^{1z}L_{i}^{2x}+L_{i}^{2z}L_{i}^{1x})-L_{i}^{2z}L_{i}^{2x}\Big\},
\end{align}
with $J_{ij}(\Bk)=\sum_{\BR}J_{\BR+\Br_i,\Br_j}e^{-i\Bk\cdot(\BR+\Br_i-\Br_j)}$, $\BD_{ij}=\sum_{\BR}\BD_{\BR+\Br_i,\Br_j}e^{-i\Bk\cdot(\BR+\Br_i-\Br_j)}$, and $\Be_{\Br}^{\pm} = \Be_{\Br}^{1} \pm i\Be_{\Br}^{2}$.

The Floquet magnon Hamiltonian of Eq.~\eqref{eq: Floquet_magnon_hamiltonian} is diagonalized by a Bogoliubov transformation that preserves the bosonic commutation relations. 
This is carried out by diagonalizing the spin wave Hamiltonian with a paraunitary matrix $T_{\Bk}$, which by definition satisfies $T_{\Bk}^\dagger \Sigma T_{\Bk}=T_{\Bk} \Sigma T_{\Bk}^\dagger=\Sigma$, with 
\begin{align}\label{eq:Sigma}
\Sigma = 
\begin{pmatrix}
\mathds{1}_{p\times p} & 0 \\
0 & - \mathds{1}_{p\times p} \\
\end{pmatrix} \,,
\end{align}
where $\mathds{1}_{p\times p}$ is the identity matrix of order $p$.
The diagonalized spin wave Hamiltonian is given by
\begin{align}
\Hsw = S\sum_{\lambda, \Bk} \CalE_\lambda(\Bk) \big(\alpha_{\Bk\lambda}^\dagger\alpha_{\Bk\lambda}+\tfrac{1}{2} \big)+\CalE_0 \,,
\end{align}
where $\lambda$ is the index for each magnon mode, $\CalE_\lambda(\Bk)$ is the corresponding energy, and $(\alpha_{\Bk},\alpha_{-\Bk}^\dagger)^T=T_{\Bk}(a_{\Bk },a_{-\Bk}^\dagger)^T$.
We use the numerical diagonalization method described in Ref.~\cite{colpaDiagonalizationQuadraticBoson1978} to obtain $T_{\Bk}$ and $\CalE_\lambda(\Bk)$. 

The Floquet magnon band topology is characterized by the Berry curvature and Chern number, which are defined for the $\lambda$-th magnon band as~\cite{shindouTopologicalChiralMagnonic2013}
\begin{align}
\Omega_\lambda&=\, i\textrm{Tr}[V_\lambda (  \partial_{k_x} T_{\Bk}^{-1} \partial_{k_y} T_{\Bk} - \partial_{k_y} T_{\Bk}^{-1} \partial_{k_x} T_{\Bk} )],\\
C_\lambda&=\int_\textrm{BZ} \frac{d\Bk}{2\pi}\Omega_\lambda,
\end{align}
where $V_\lambda$ is a $2p\times 2p$ matrix whose matrix element is unity only at the $\lambda$-th diagonal element and zero otherwise.

\begin{figure}[t]
    \centering
    \includegraphics[width=\columnwidth]{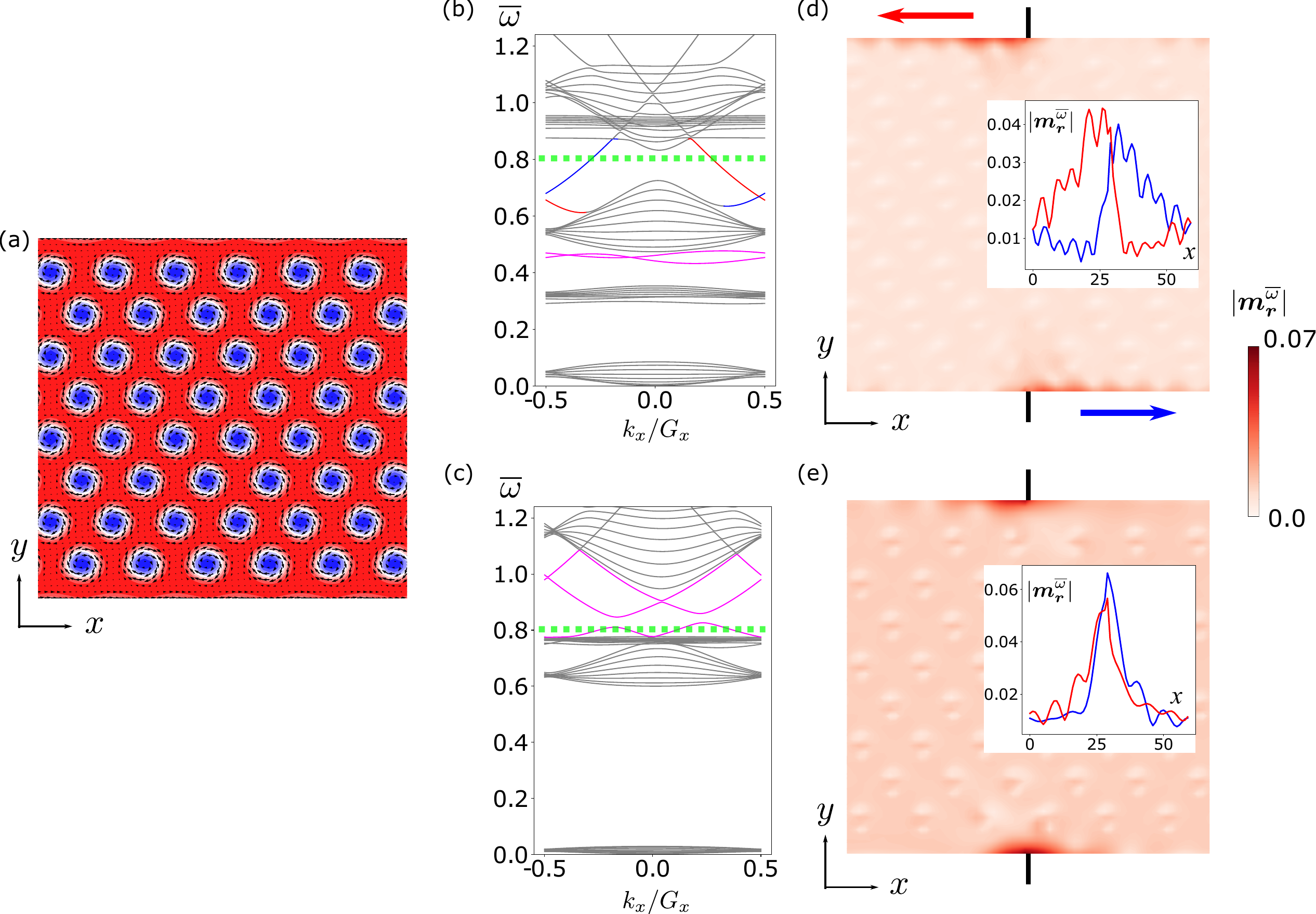}
    \caption{\textbf{Ultrafast switching of chiral magnonic edge states.} (a) Classical ground-state spin configuration in a strip geometry at $D/J=1.0$ and $\overline{B}_0=0.4$. (b)-(c) Floquet magnon edge spectra under RCP light at (b) $\tilde{B}_d=1.5$ and (c) $\tilde{B}_d=2.7$. The other parameters are fixed at $\tilde{\omega}_0=10$, $\alpha=0.04$, and $\CalE_d=0.1$. $G_s$ denotes the primitive reciprocal lattice vector amplitude. (d)-(e) Steady-state amplitude of oscillations $| \Bm_{\Br}^{\overline{\omega}}|$, defined in Eq.~\eqref{eq: amplitude_filtered_SW}, obtained for the same parameters as in (b) and (c), respectively.
    Here, $| \Bm_{\Br}^{\overline{\omega}}|$ is obtained under an additional local in-plane ac magnetic field, applied at single edge sites indicated by black vertical lines, whose frequency $\overline{\omega}=0.8$ lies inside the band gap [green dotted lines in (b) and (c)]. Insets in (d) and (e) show $| \Bm_{\Br}^{\overline{\omega}}|$ along the top (red) and bottom (blue) edge of the strip, with chiral propagation of magnonic edge states emerging only at $\tilde{B}_d=1.5$.
     }
    \label{fig: edge_LLG_magnon}
\end{figure}
\section{Laser-controlled chiral magnonic edge states}
\label{section: edge states}

In the main text, we have demonstrated a laser-induced topological phase transition in the bulk Floquet magnon spectrum at $\tilde{B}_d \approx 1.8$.
To confirm the bulk-boundary correspondence, in this section we study the emergence of chiral magnonic edge states under laser irradiation.
Figure~\ref{fig: edge_LLG_magnon}(a) shows a skyrmion crystal formed in a strip geometry, with periodic boundary conditions along the $x$-axis and open boundary conditions along the $y$-axis, obtained by Monte Carlo annealing at $D/J=1.0$ and $\overline{B}_0=0.4$. 

\begin{figure}[t]
    \centering
    \includegraphics[width=\columnwidth]{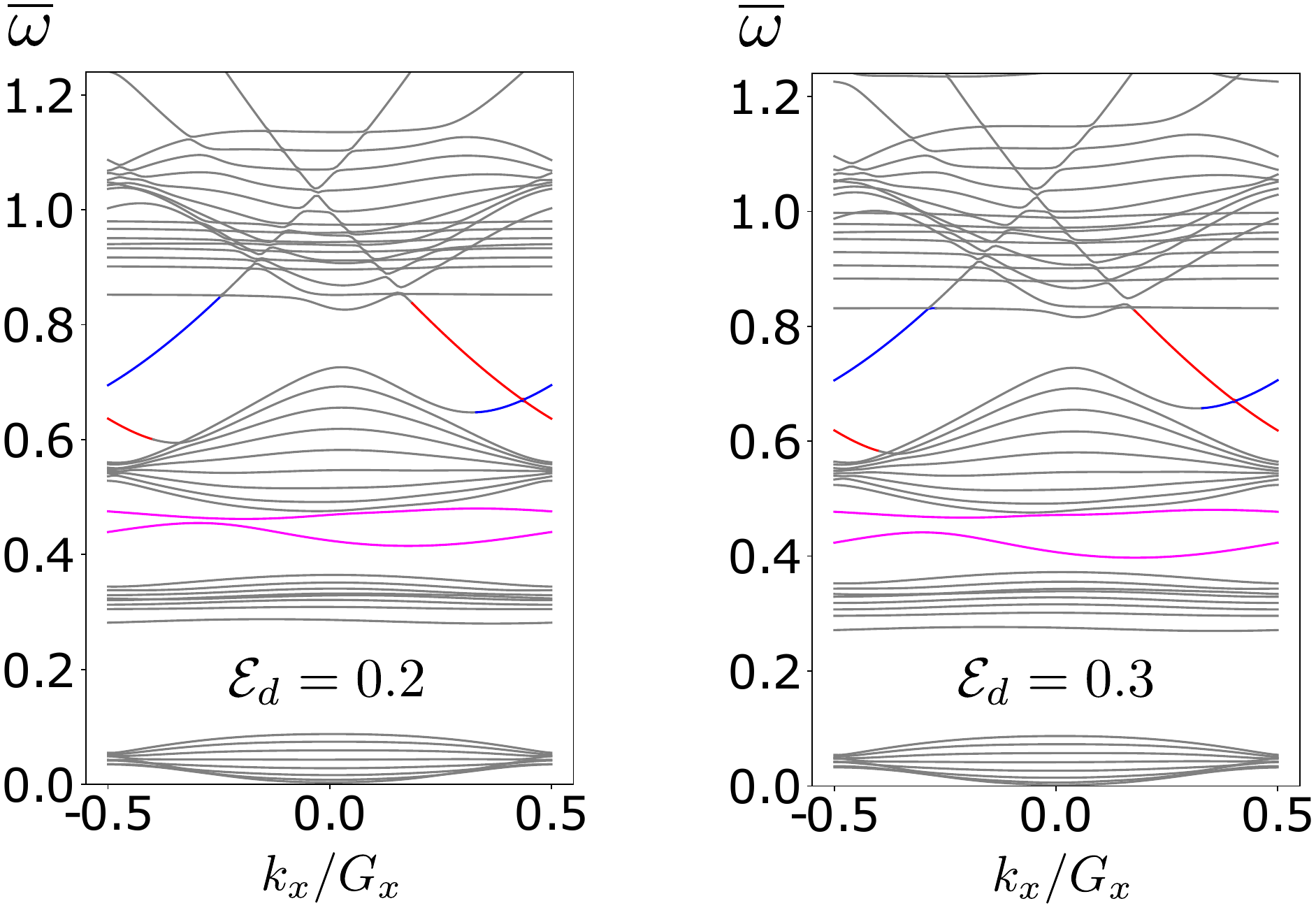}
    \caption{\textbf{Nonreciprocal magnonic edge states due to the magnetoelectric coupling.} Floquet magnon edge spectra under RCP light at $\CalE_d=0.2$ (left panel) and $\CalE_d=0.3$ (right panel). The other parameters are the same as  in Fig.~\ref{fig: edge_LLG_magnon}(b).
     }
    \label{fig: non_reciprocity}
\end{figure}

The NESS of the skyrmion crystal under RCP light is obtained by solving the effective LLG equation for $\textbf{B}_0\parallel[110]$.
The parameters are fixed at $\CalE_d=0.1$, $\alpha=0.04$, and $\tilde{\omega}_0=10$.
Using Eq.~\eqref{eq: Floquet_magnon_hamiltonian}, the edge spectra of Floquet magnons at $\tilde{B}_d=1.5$ and $\tilde{B}_d=2.7$ are computed from the time-averaged spin configuration of the NESS, as shown in Figs.~\ref{fig: edge_LLG_magnon}(b)-(c).
Since the total Chern number below the third band gap, between the CCW and breathing modes, is +1 at $\tilde{B}_d=1.5$ (see Fig.~3 of the main text), we expect the presence of a topologically protected edge state within this gap.
This is illustrated in Fig.~\ref{fig: edge_LLG_magnon}(b), where the right-moving and left-moving edge states are highlighted in blue and red, respectively. We note that the band crossing between the chiral edge states is slightly shifted from $k_x/G_x=0.5$, indicating the nonreciprocal nature of magnon bands due to the asymmetry between $+k_x$ and $-k_x$. The nonreciprocity becomes more pronounced for larger $\CalE_d$ as shown in Fig.~\ref{fig: non_reciprocity}. In addition, there exist topologically trivial edge states highlighted in magenta. 
On the other hand, as illustrated in Fig.~\ref{fig: edge_LLG_magnon}(c), the edge spectrum of the topologically trivial phase at $\tilde{B}_d=2.7$ no longer features chiral edge states within the third band gap, while trivial edge states remain.

To reveal the chirality of the edge states, we compute the LLG equation under RCP light with the same parameters as in Figs.~\ref{fig: edge_LLG_magnon}(b)-(c). The chiral magnonic edge states are excited by additionally applying a local in-plane ac magnetic field at a single site at the top and bottom edges of the sample, given by $g\mu_B\textbf{B}_\textrm{ac}(t)/JS=0.1\cos(\omega t)\hat{y}$ with $\overline{\omega}=0.8$.
After $\overline{t}=$10,000, we obtain the time-periodic steady state with edge-localized spin waves emitted from the sites excited by the additional field.
To filter out oscillations at $\tilde{\omega}_0$ induced by the laser, we take the Fourier transform of the magnetization at each site to extract the Fourier component at resonance with $\textbf{B}_\textrm{ac}(t)$, denoted as
\begin{equation}
    \Bm_{\Br}^{\overline{\omega}}=\Bm^{\overline{\omega}}_{\Br,+1}e^{-i\omega t}+\Bm^{\overline{\omega}}_{\Br,-1}e^{i\omega t},
    \label{eq: amplitude_filtered_SW}
\end{equation}
with $\overline{\omega}=0.8$ and where $\Bm^{\overline{\omega}}_{\Br,\pm1}$ is a three-dimensional vector with complex values. We should note that $|\Bm_{\Br}^{\overline{\omega}}|$ is not normalized to unity. 

Figures~\ref{fig: edge_LLG_magnon}(d)-(e) show the amplitude of oscillations at resonance with $\textbf{B}_\textrm{ac}(t)$, namely $| \Bm_{\Br}^{\overline{\omega}}|$.
In Fig.~\ref{fig: edge_LLG_magnon}(d) at $\tilde{B}_d=1.5$, we find chiral spin waves propagating to the left (right) along the top (bottom) edge of the sample.
This is the manifestation of chiral magnonic edge states.
The chiral propagation of the edge states is further confirmed in the inset, where we plot $| \Bm_{\Br}^{\overline{\omega}}|$ as a function of position along the $x$-axis at the top (red) and bottom (blue) edge. 
In contrast, there is no signature of edge-localized spin wave propagation at $\tilde{B}_d=2.7$ due to the presence of a global band gap in the topologically trivial phase, as shown in Fig.~\ref{fig: edge_LLG_magnon}(e). 
In addition, the inset shows a symmetric and rapid decay of $| \Bm_{\Br}^{\overline{\omega}}|$.
Therefore, our results demonstrate the ultrafast switching of topologically protected chiral magnonic edge states using circularly polarized lasers.

\section{Floquet magnonic thermal Hall conductivity}

When the CCW/breathing mode is resonantly excited, the distribution function of the $n$-th magnon band can be approximated in the thermodynamic limit as~\cite{nakataMagnonTransportMicrowave2015} 
$\rho_{n,\Bk}\approx\frac{|I|^2}{(\omega_{c/b}-\omega_{n,\Bk})^2+\alpha_{\textrm{eff},c/b}^2\omega_{c/b}^2}\delta(\Bk)\delta_{n,c/b}
$, where $c/b$ corresponds to the band index of the CCW/breathing mode with resonance frequency $\omega_{c/b}$, $\alpha_{\textrm{eff},c/b}$ is the effective damping constant for each mode, and $I=g\mu_BB_\textrm{pump}/\hbar$ with $B_\textrm{pump}$ the amplitude of the pumping field at frequency $\omega_{c/b}$. 
Here, the resonance frequencies $\omega_{c/b}$ are computed as functions of the laser field $B_d$.
The effective damping constant is given by
$
\alpha_{\textrm{eff},c/b}=1/(\tau_{c/b}\omega_{c/b})$ with $\tau_{c/b}$ denoting the lifetime of each spin wave mode~\cite{rozsaEffectiveDampingEnhancement2018}.
From the LLG simulation, we estimate $\alpha_{\textrm{eff},c}=0.0985$ and $\alpha_{\textrm{eff},b}=0.0548$. We should note that they are enhanced from the Gilbert damping constant $\alpha=0.04$ due to the noncollinear spin configuration~\cite{rozsaEffectiveDampingEnhancement2018}.

Furthermore, we need to account for a broadening in crystal momentum space due to finite size effects.
This is achieved by replacing $\delta(\Bk)$ with a Gaussian distribution. The final expression of the distribution function is thus given by
\begin{align}
    \rho_{n,\Bk}&\approx\frac{|I|^2\delta_{n,c/b}}{(\omega_{c/b}-\omega_{n,\Bk})^2+\alpha^2_{\textrm{eff},c/b}\omega_{c/b}^2}\nonumber\\
    &\times\frac{N^2\exp\Big[-\frac{N^2}{2}\big(\frac{ k_x^2}{G_x^2}+\frac{ k_y^2}{G_y^2}\big)\Big]}{2\pi }\,,
\end{align}
with $N^2$ denoting the number of magnetic unit cells in the sample, $G_x$ and $G_y$ denoting reciprocal lattice vectors. 
Substituting $\rho_{n,\boldsymbol{k}}$ into $c_2(\rho)$, we find that $c_2(\rho_{n,\boldsymbol{k}})\approx \frac{\pi^2H(|\boldsymbol{k}|-G_1)}{3}\delta_{n,c/b}$ provided $|I|^2\sim \alpha^2_{\textrm{eff},c/b}\omega_{c/b}^2$, where $H(x)$ denotes the Heaviside step function and $G_1$ depends on $N$ and is numerically determined.
In the asymptotic limit of $N \rightarrow \infty$, the thermal Hall conductivity vanishes with $G_1\rightarrow 0$. 
However, the Berry curvature diverges at the $\Gamma$ point when the gap between the CCW and breathing modes closes, thus inducing a measurable magnonic thermal Hall effect even for a large $N$.
This is demonstrated in Fig.~4 of the main text for $N=100$ and $I=0.1$ (scaled to $7$ mT from Table~1 in the main text), obtained by the numerical integration of 200$\times$200 mesh points over $\pm 0.2 G_x$ and $\pm 0.2 G_y$.

\section{Laser parameters and heating effect}
\label{section: heating}
Using Table~1 in the main text, we estimate the laser frequency to be greater than 1.25 THz for $\tilde{\omega}_0\ge 5$ and the amplitude to be $1.6$ T (4.8 MV/cm) at $\tilde{B}_d\sim 2.0$.
Noting that the electric field strength could exceed 10 MV/cm in strong THz laser sources~\cite{pashkinQuantumPhysicsUltrabroadband2013, fulopLaserDrivenStrongFieldTerahertz2020}, our theory can be experimentally realized with the already available equipment. 

Another issue for the experimental realization is the heating effect.
We note that the direct absorption of photons is not possible for THz lasers as the energy of photons is much smaller than the electronic bulk band gap of Cu$_2$OSeO$_3$, $\Delta_g\approx 2 $ eV~\cite{versteegOpticallyProbedSymmetry2016}. 
Thus, the spontaneous heating of electrons should not occur~\cite{ogawaUltrafastOpticalExcitation2015a}.
Nevertheless, we cannot completely suppress the heating effect that could eventually annihilate the long-range magnetic order.
To circumvent this issue, we suggest repeated applications of ultrashort pulses of circularly polarized laser as in the case of light-induced anomalous Hall effect in graphene~\cite{mciverLightinducedAnomalousHall2020, okaFloquetEngineeringQuantum2019}.

\begin{figure}
    \centering
    \includegraphics[width=\columnwidth]{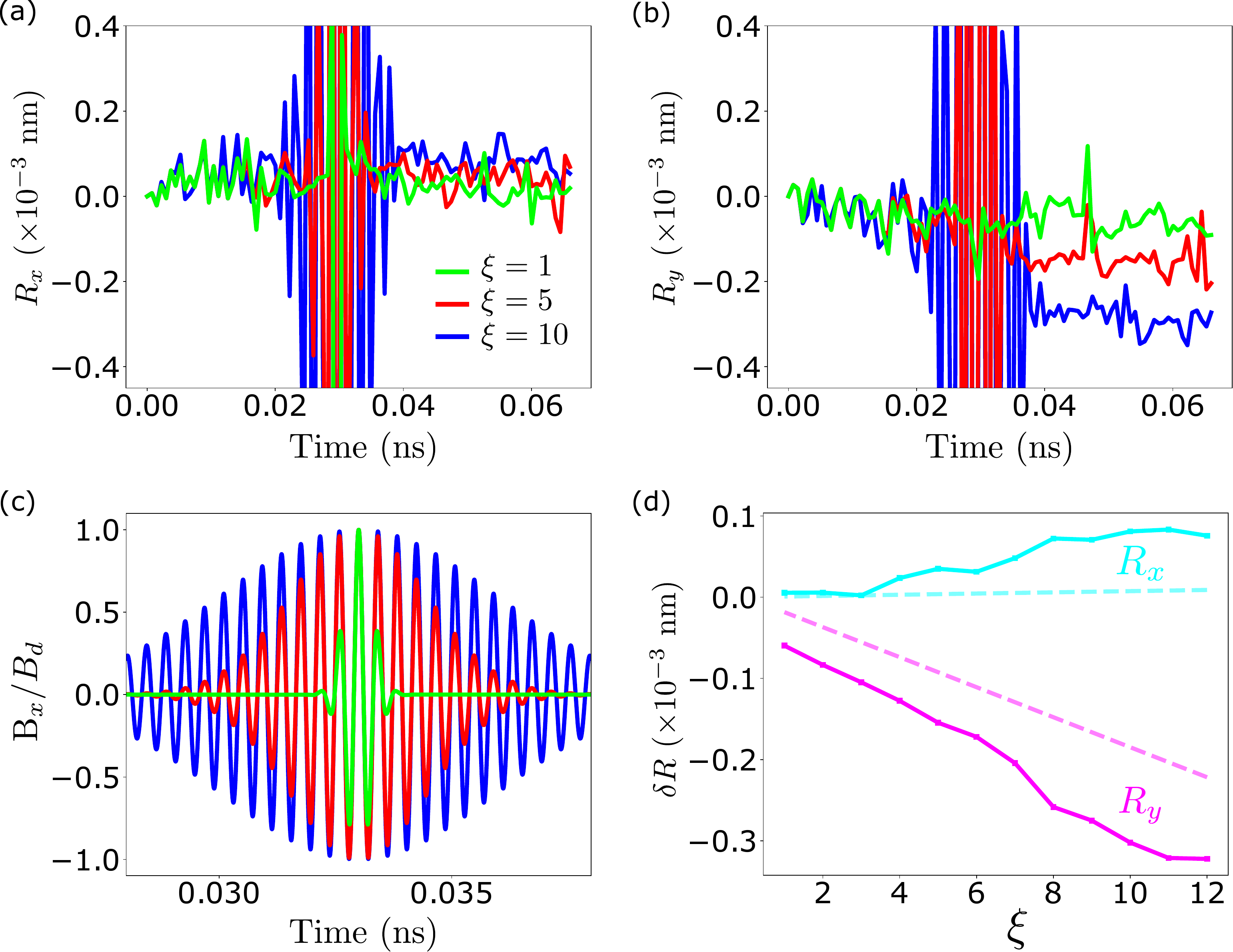}
    \caption{\textbf{Skyrmion motion induced by an ultrashort laser pulse.} (a),(b)~Displacement of the center of a skyrmion (a) $R_x$ and (b) $R_y$ under an LCP laser pulse for $\textbf{B}_0\parallel [110]$. The parameters are taken to be the same as in Fig.~2 of the main text, where $D/J=0.09$, $\overline{B}_0=0.9$, $\tilde{B}_d=2.0$, $\tilde{\omega}_0=10$, $\CalE_d=0.1$, and $\alpha=0.04$. (c) Magnetic field along $x$-axis under the laser pulse. The pulse field is described by a Gaussian envelop function with parameters $\xi$ and $t_0$, defined in Eq.~\eqref{eq: Gaussian_pulse}. In (a)-(c), each plot respectively corresponds to $\xi=1$ (lime), $\xi=5$ (red), and $\xi=10$ (blue) while we fix the value of $t_0$ as 0.33~ns. (d) Total displacement of the skyrmion center $R_x$ (cyan) and $R_y$ (magenta) as a function of $\xi$ for $\textbf{B}_0\parallel [110]$. The other parameters are the same as in (a)-(c). Dashed lines are obtained by $\delta R_{x/y}=v_{x/y}\delta T$ using Eq.~\eqref{eq: SM_velocity_formula} and the duration of the pulse $\delta T\approx 2\xi T_0$.
     }
    \label{fig: pulse_response}
\end{figure}

\begin{figure}
    \centering
    \includegraphics[width=\columnwidth]{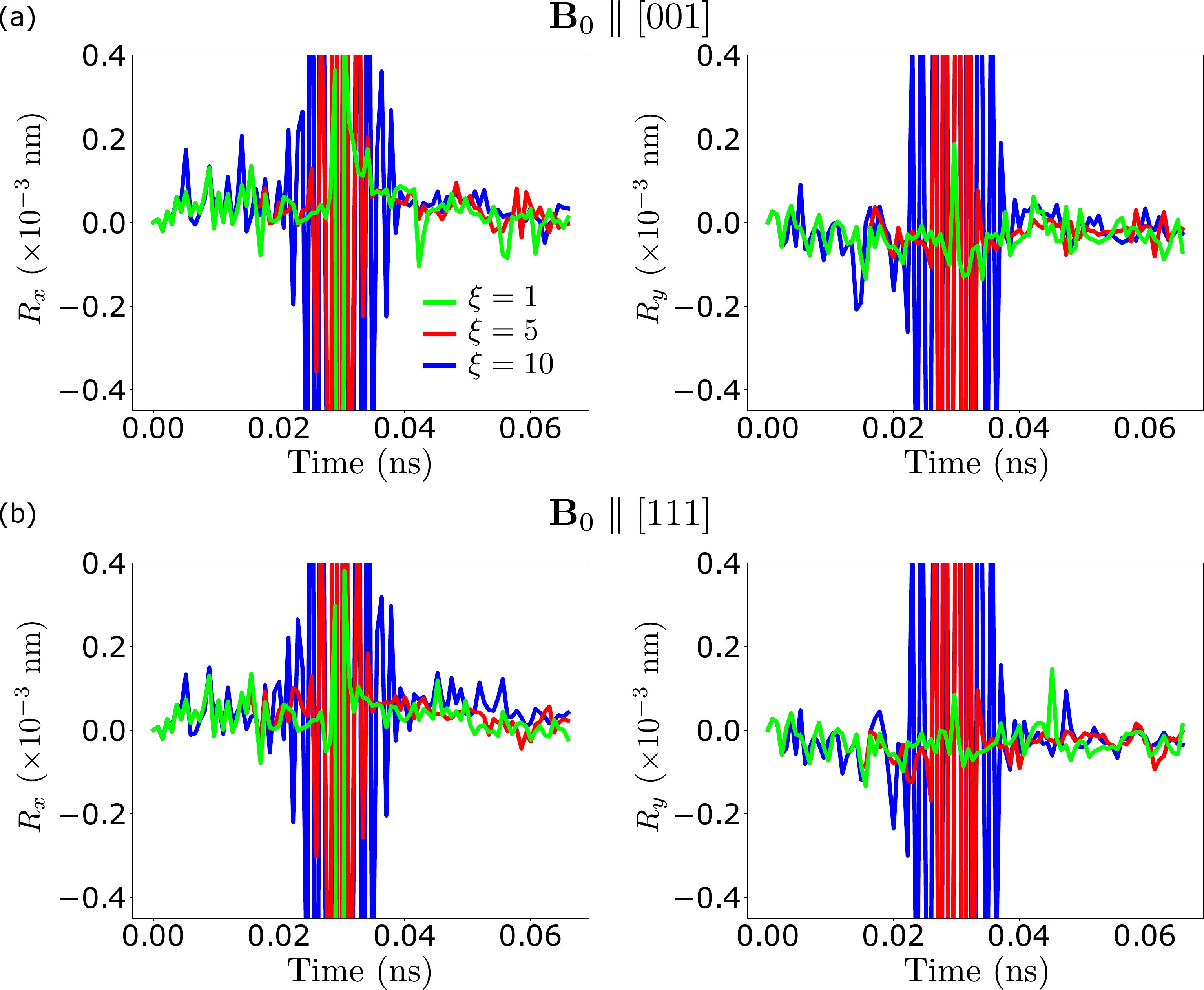}
    \caption{\textbf{No skyrmion motion induced by an ultrashort laser pulse for $\textbf{B}_0\parallel [001]$ and $\textbf{B}_0\parallel [111]$.} (a),(b)~Displacement of the center of a skyrmion $R_x$ (left) and $R_y$ (right) under an LCP laser pulse for (a) $\textbf{B}_0\parallel [001]$ and (b) $\textbf{B}_0\parallel [111]$. The parameters are taken to be the same as in Fig.~\ref{fig: pulse_response}(a)-(c).
     }
    \label{fig: pulse_response_other}
\end{figure}
\section{Skyrmion motion under ultrashort laser pulses}
In Sec.~\ref{section: heating}, we suggested using ultrashort pulses to avoid heating effects. It should be noted that our results cannot be applied directly to skyrmions under laser pulses, since they are no longer driven by time-periodic drives with a single frequency~\cite{toyotaEnvelopeHamiltonianElectron2015, medisauskasFloquetHamiltonianApproach2018}. To account for this difference, we numerically study the dynamics of single skyrmions under ultrashort pulses. 
Time-dependent electromagnetic fields of ultrashort pulses are described as $\textbf{B}'(t)=\textbf{B}(t)A(t)$ and $\textbf{E}'(t)=\textbf{E}(t)A(t)$, where the Gaussian envelop function is defined as
\begin{equation}
A(t)=\exp\Big[-\Big(\frac{t-t_0}{\xi T_0}\Big)^2\Big],
\label{eq: Gaussian_pulse}
\end{equation}
with $T_0=2\pi/\omega_0$ and parameters $\xi$ and $t_0$. With a larger value of $\xi$, the number of oscillations inside a single pulse increases [see Fig.~\ref{fig: pulse_response}(c)]. In the limit of $\xi\rightarrow \infty$, we recover the situation where the system is under a time-periodic drive that is well described by Floquet theory. Remarkably, we find that skyrmion dynamics under ultrashort pulses is in agreement with Floquet theory even for small values of $\xi$ as discussed below.

In the following, we reduce the discretized time step for the LLG equation solver to $\Delta \overline{t}=10^{-3}$ in order to simulate skyrmion dynamics within an ultrashort time scale. In Fig.~\ref{fig: pulse_response}(a) and (b), we show the change in the skyrmion center coordinates, $R_x$ and $R_y$, for $\textbf{B}_0\parallel [110]$ under irradiation of an LCP pulse with $\xi=1$ (lime), $\xi=5$ (red), and $\xi=10$ (blue), respectively.  It clearly shows that skyrmions are driven along the $-y$ direction similar to the laser-induced motion shown in Fig.~2 of the main text. In addition, the amplitude of the displacement becomes larger as we increase the value of $\xi$ as shown in Fig.~\ref{fig: pulse_response}(b). Estimating the final position of the skyrmion center by averaging $R_x$ and $R_y$ over 0.01~ns, the total displacement$~\delta R$ from the initial position is plotted for  various values of $\xi$ in Fig.~\ref{fig: pulse_response}(d). We find that the total displacement is linearly proportional to $\xi$ with a positive/negative gradient for $R_x/R_y$, which is consistent with Eq.~\eqref{eq: SM_velocity_formula}. Furthermore, by estimating the duration of the pulse as $\delta T\approx 2\xi T_0$, we obtain  good agreement with Eq.~\eqref{eq: SM_velocity_formula} for $R_y$, although the numerical error in $R_x$ and $R_y$ results in some deviations. Therefore, the total displacement of skyrmions under laser pulses can be estimated from the duration of the pulses using Eq.~\eqref{eq: SM_velocity_formula}.

The above numerical study shows that our results based on Floquet theory are both qualitatively and quantitatively consistent with skyrmion motion under ultrashort pulses. Here, we provide a phenomenological argument to explain why this could be the case, although a more detailed study is necessary in the future.
When the pulse field is applied, we find forced oscillations in $R_x$ and $R_y$ as shown in Fig.~\ref{fig: pulse_response}(a) and (b). During each oscillation, with period $T_0$, skyrmions experience a nonzero net force due to the magnetoelectric coupling, which approaches Eq.~\eqref{eq: F_110} asymptotically for larger $\xi$. Crucially, even under ultrashort pulses with small $\xi$, the direction of the net force exerted on skymions seems to be the same as in Eq.~\eqref{eq: F_110}. We argue that this is because the net force is determined by the symmetry of the magnetoelectric coupling. Indeed, as shown in Fig.~\ref{fig: pulse_response_other}, no translational motion is obtained under ultrashort pulses for $\textbf{B}_0\parallel [001]$ or $\textbf{B}_0\parallel [111]$, regardless of the value of $\xi$. Therefore, skyrmions can be driven controllably under ultrashort laser pulses similarly to time-periodic electromagnetic fields.